\documentclass[aps,preprint,groupedaddress,showpacs]{revtex4}
\usepackage{graphicx}
\begin{document}

\title{The analytical $\pi\pi$ scattering amplitude and the light scalars}
\author {
N.N. Achasov$^{\,a}$ \email{achasov@math.nsc.ru} and A.V.
Kiselev$^{\,a,b}$ \email{kiselev@math.nsc.ru}}

\affiliation{
   $^a$Laboratory of Theoretical Physics,
 Sobolev Institute for Mathematics, 630090, Novosibirsk, Russia\\
$^b$Novosibirsk State University, 630090, Novosibirsk, Russia}

\date{\today}

\begin{abstract}

In this work we construct the $\pi\pi$ scattering amplitude
$T^0_0$ with regular analytical properties in the $s$ complex
plane, which describes simultaneously the data on the $\pi\pi$
scattering, $\phi\to\pi^0\pi^0\gamma$ decay, and $\pi\pi\to K\bar
K$ reaction. The chiral shielding of the $\sigma (600)$ meson and
its mixing with the $f_0(980)$ meson are also taken into account.
The data agrees with the four-quark nature of the $\sigma (600)$
and $f_0(980)$ mesons. The amplitude in the range $-5 m_\pi^2 < s
< 0.64$ GeV$^2$ also agrees with results, obtained on the base of
the chiral expansion, dispersion relations, and the Roy equations.
\end{abstract}

\pacs{12.39.-x  13.40.Hq  13.66.Bc}

\maketitle

\section{Introduction}

Study of light scalar resonances is one of the central problems of
nonperturbative QCD, it is important for understanding both the
confinement physics and the chiral symmetry realization way in the
low energy region. The commonly suggested nonet of light scalar
mesons is $f_0(600)$ [or $\sigma (600)$], $K_0^*(800)$ [or $\kappa
(800)$], $f_0(980)$, and $a_0(980)$ \cite{pdg-2008}. Light scalar
mesons are intensively studied theoretically and experimentally in
different reactions.

In Refs. \cite{our_f0} we described the high-statistical KLOE data
on the $\phi\to\pi^0\pi^0\gamma$ decay \cite{pi0publ}
simultaneously with the data on the $\pi\pi$ scattering and the
$\pi\pi\to K\bar K$ reaction. The description was carried out
taking into account the chiral shielding of the $\sigma (600)$
meson \cite{annshgn-94,annshgn-07} and its mixing with the
$f_0(980)$ meson. It was shown that the data do not contradict the
existence of the $\sigma (600)$ meson and yield evidence in favor
of the four-quark nature of the $\sigma (600)$ and $f_0(980)$
mesons.

This description revealed new goals. The point is that at the same
time it was calculated in Ref. \cite{sigmaPole} the $\pi\pi$
scattering amplitude in the $s$ complex plane, basing on chiral
expansion, dispersion relations, and Roy equations. In particular,
the pole was obtained at $s=M_\sigma^2=(6.2-12.3i)\,m_\pi^2$,
where

\begin{equation}
M_\sigma = 441^{+16}_{-8} - i272^{+9}_{-12.5}\ \mbox{MeV}\,,
\label{poleSigma}\end{equation}

\noindent which was assigned to the $\sigma$ resonance.

Aiming the comparison of the results of Refs. \cite{our_f0} and
\cite{sigmaPole} it is necessary to build the $\pi\pi$ scattering
amplitude with correct analytical properties in the complex $s$
plane. The point is that in Ref. \cite{our_f0} S matrix of the
$\pi\pi$ scattering is the product of the "resonance" and
"background" parts:

\begin{equation}
S_{\pi\pi} = S_{back}\,S_{res}\,, \label{SSrepres}
\end{equation}

\noindent and the $S_{res}$ had correct analytical properties,
while analytical properties of the $S_{back}$ in the whole complex
$s$ plane were not essential for the aims of \cite{our_f0}, where
the physical region was investigated, and Adler zero existence
\cite{adler-65b} together with the poles absence on the real axis
of the $s$ complex plane were demanded.

In this paper we present the $\pi\pi$ scattering amplitude with
correct analytical properties in the complex $s$ plane and the
data description obtained with this amplitude \cite{TbackIs0}. The
comparison with the results of Ref. \cite{sigmaPole} is also
presented.

All formulas for the $\phi\to(S\gamma+\rho^0\pi^0)\to\pi^0
\pi^0\gamma$ reaction [$S=f_0(980)+\sigma(600)$] are shown in Sec.
\ref{sform}. Our new parametrization of the background amplitude
is presented in Secs. \ref{sBackPhase} and \ref{restr}. The
results of the data analysis are presented in Sec. \ref{sr}. A
brief summary is given in Sec. \ref{sc}.

\section{Theoretical description of the \lowercase{$\phi\to
(f_0(980)+\sigma(600))\gamma\to\gamma\pi^0\pi^0$} and
\lowercase{$\phi\to\rho^0\pi^0\to\gamma\pi^0\pi^0$} reactions }
\label{sform}

In Refs. \cite{achasov-89,achasov-97} it was shown that the
dominant background process is
$\phi\to\pi^0\rho\to\gamma\pi^0\pi^0$, while the reactions
$e^+e^-\to\rho\to\pi^0\omega\to\gamma\pi^0\pi^0$ and
$e^+e^-\to\omega\to\pi^0\rho\to\gamma\pi^0\pi^0$ have a small
effect on $e^+e^-\to\phi\to\gamma\pi^0\pi^0$ in the region
$m_{\pi^0\pi^0}\equiv m>900$ MeV. In Ref. \cite{a0f0} it was shown
that the $\phi\to\pi^0\rho\to\gamma\pi^0\pi^0$ background is small
in comparison with the signal $\phi\to\gamma
f_0(980)\to\gamma\pi^0\pi^0$ at $m>700$ MeV.

The amplitude of the background decay
$\phi(p)\to\pi^0\rho\to\gamma(q)\pi^0(k_1)\pi^0(k_2)$ has the
following form:

\begin{equation}
M_{back}=F_b
e^{-i\delta}g_{\rho\pi^0\phi}g_{\rho\pi^0\gamma}\phi_{\alpha}p_{\nu}
\epsilon_{\delta}q_{\epsilon}\epsilon_{\alpha\beta\mu\nu}
\epsilon_{\beta\delta\omega\epsilon}\bigg(\frac{k_{1\mu}k_{2\omega}}
{D_{\rho}(q+k_2)}+\frac{k_{2\mu}k_{1\omega}}{D_{\rho}(q+k_1)}\bigg).
\label{amp_back}
\end{equation}

Here, constants $F_b$ and $\delta$ take into account $\rho\pi$
rescattering effects \cite{rhophase}. Note that in this work and
our previous work it was assumed that $F_b=1$ \cite{smallRhoPi}.

In the $K^+K^-$ loop model, $\phi\to K^+K^-\to \gamma(f_0+\sigma)$
\cite{achasov-89,achasov-97,a0f0}, above the $K\bar K$ threshold
the amplitude of the signal
$\phi\to\gamma(f_0+\sigma)\to\gamma\pi^0\pi^0$  is
\begin{equation}
M_{sig}=g(m)\bigg((\phi\epsilon)- \frac{(\phi q)(\epsilon
p)}{(pq)}\bigg)\,T\left(K^+K^-\to\pi^0\pi^0\right )\times 16\pi
\label{f0signal}\,,
\end{equation}
where the $K^+K^-\to\pi^0\pi^0$ amplitude, taking into account the
mixing of $f_0$ and $\sigma$ mesons,
\begin{equation}
T\left(K^+K^-\to\pi^0\pi^0\right ) =
e^{i\delta_B}\sum_{R,R'}\frac{g_{RK^+K^-}G_{RR'}^{-1}g_{R'\pi^0\pi^0}}{16\pi}\,,
\label{kktopipi}
\end{equation}
where $R,R'=f_0,\sigma$,
\begin{equation}
\delta_B =\delta^{\pi\pi}_B+\delta^{K\bar K}_B\,, \label{delph}
\end{equation}
where $\delta^{\pi\pi}_B$ and $\delta_B^{K\bar K}$ are phases of
the elastic background of the $\pi\pi$ and $K\bar K$ scattering,
respectively, see Refs. \cite{adsh-80,adsh-1980,z_phys,adsh-84}.

Note that the additional phase $\delta_B^{K\bar K}$ changes the
modulus of the $K\bar K\to\pi^0\pi^0$ amplitude under the $K\bar
K$ threshold, at $m<2 m_{K}$. Let us define

\[P_K= \left\{\begin{array}{ll}
  e^{i\delta_B^{K\bar K}}\hspace{60 mm} m\geq 2
m_{K}\,;\\
 \mbox{analytical continuation of } e^{i\delta_B^{K\bar
 K}}\hspace{12 mm}
m<2 m_{K}\,.\hspace{53 mm}\addtocounter{equation}{1}
(\theequation)
 \label{Kphas}
\end{array}\right.\]

Note also that the phase $\delta_B^{\pi\pi}$ was defined as
$\delta_B$ in Refs. \cite{achasov-97,a0f0}.

The matrix of the inverse propagators \cite{achasov-97} is

\[G_{RR'}\equiv G_{RR'}(m)=\left( \begin{array}{cc} D_{f_0}(m)&-\Pi_{f_0\sigma}(m)\\-\Pi_{f_0 \sigma}(m)&D_{\sigma}(m)\end{array}\right),\]

$$\Pi_{f_0 \sigma}(m)=\sum_{a,b} \frac{g_{\sigma ab}}{g_{f_0
ab}}\Pi^{ab}_{f_0}(m)+C_{f_0\sigma},$$

\noindent where the constant $C_{f_0\sigma}$ incorporates the
subtraction constant for the transition $f_0(980)\to(0^-0^-)\to
\sigma (600)$ and effectively takes into account the contribution
of multiparticle intermediate states to $f_0\leftrightarrow\sigma$
transition, see Ref. \cite{achasov-97}. The inverse propagator of
the R scalar meson is also presented in Refs.
\cite{achasov-89,achasov-97,a0f0,adsh-79,adsh-80,adsh-1980,z_phys,adsh-84,ourProp,achasov-95,achasov-84,achasov-01a,our_a0}:

\begin{equation}
\label{propagator} D_R(m)=m_R^2-m^2+\sum_{ab}[Re
\Pi_R^{ab}(m_R^2)-\Pi_R^{ab}(m^2)],
\end{equation}
where $\sum_{ab}[Re \Pi_R^{ab}(m_R^2)-
\Pi_R^{ab}(m^2)]=Re\Pi_R(m_R^2)- \Pi_R(m^2)$ takes into account
the finite width corrections of the resonance which are the one
loop contribution to the self-energy of the $R$ resonance from the
two-particle intermediate  $ab$ states.

For pseudoscalar $a,b$ mesons and $m_a\geq m_b,\ m\geq m_+$ one
has

\begin{eqnarray}
\label{polarisator}
&&\Pi^{ab}_R(m^2)=\frac{g^2_{Rab}}{16\pi}\left[\frac{m_+m_-}{\pi
m^2}\ln \frac{m_b}{m_a}+\right.\nonumber\\
&&\left.+\rho_{ab}\left(i+\frac{1}{\pi}\ln\frac{\sqrt{m^2-m_-^2}-
\sqrt{m^2-m_+^2}}{\sqrt{m^2-m_-^2}+\sqrt{m^2-m_+^2}}\right)\right]
\end{eqnarray}
При $m_-\leq m<m_+$
\begin{eqnarray}
&&\Pi^{ab}_{R}(m^2)=\frac{g^2_{Rab}}{16\pi}\left[\frac{m_+m_-}{\pi
m^2}\ln \frac{m_b}{m_a}-|\rho_{ab}(m)|+\right.\nonumber\\
&&\left.+\frac{2}{\pi}|\rho_{ab}(m)
|\arctan\frac{\sqrt{m_+^2-m^2}}{\sqrt{m^2-m_-^2}}\right].
\end{eqnarray}
При $m<m_-$
\begin{eqnarray}
&&\Pi^{ab}_{R}(m^2)=\frac{g^2_{Rab}}{16\pi}\left[\frac{m_+m_-}{\pi
m^2}\ln \frac{m_b}{m_a}-\right.\nonumber\\
&&\left.-\frac{1}{\pi}\rho_{ab}(m)\ln\frac{\sqrt{m_+^2-m^2}-
\sqrt{m_-^2-m^2}}{\sqrt{m_+^2-m^2}+\sqrt{m_-^2-m^2}}\right].
\end{eqnarray}
\noindent и
\begin{equation}
\label{rho-ab}
\rho_{ab}(m)=\sqrt{(1-\frac{m_+^2}{m^2})(1-\frac{m_-^2}{m^2})}\,\,,\qquad
m_+=m_a\pm m_b
\end{equation}

The constants  $g_{Rab}$ are related to the width
\begin{equation}
\Gamma_R(m)=\sum_{ab} \Gamma(R\to
ab,m)=\sum_{ab}\frac{g_{Rab}^2}{16\pi m}\rho_{ab}(m).
\label{f0pipi}
\end{equation}

Note that we take into account intermediate states $\pi\pi,K\bar
K,\eta\eta,\eta '\eta,\eta'\eta'$ in the $f_0(980)$ and $\sigma
(600)$ propagators:

\begin{equation}
\Pi_{f_0}=\Pi_{f_0}^{\pi^+\pi^-}+\Pi_{f_0}^{\pi^0\pi^0}+\Pi_{f_0}^{K^+K^-}+
\Pi_{f_0}^{K^0\bar{K^0}}+\Pi_{f_0}^{\eta \eta}+\Pi_{f_0}^{\eta '
\eta}+\Pi_{f_0}^{\eta ' \eta '} ,
\end{equation}

\noindent and also for the $\sigma (600)$. We use
$g_{f_0K^0\bar{K^0}}=g_{f_0K^+K^-},
g_{f_0\pi^0\pi^0}=g_{f_0\pi^+\pi^-}/\sqrt{2}$, the same for the
$\sigma (600)$, too.

For other coupling constants the naive four-quark model predicts
\cite{achasov-89,achasov-84}

$$g_{f_0\eta \eta}=-g_{f_0\eta '\eta '}=\frac{2\sqrt{2}}{3}\,
g_{f_0K^+K^-},\,\,g_{f_0\eta
'\eta}=-\frac{\sqrt{2}}{3}\,g_{f_0K^+K^-}\,;$$

$$g_{\sigma\eta \eta}=g_{\sigma\eta \eta '}=\frac{\sqrt{2}}{3}\,
g_{\sigma \pi^+\pi^-},\,\,g_{\sigma\eta '\eta
'}=\frac{1}{3\sqrt{2}}\,g_{\sigma \pi^+\pi^-}\,.$$

The definition of $g_{R\pi^0\pi^0}$, $g_{R\eta\eta}$,
$g_{R\eta'\eta'}$ takes into account the identity of the
particles. As these relations are approximate, we introduce the
effective correction coefficients $x_\sigma$ and $x_{f_0}$:

$$g_{f_0\eta \eta}=-g_{f_0\eta '\eta '}=\frac{2\sqrt{2}}{3}\,
g_{f_0K^+K^-}\,x_{f_0},\,\,g_{f_0\eta
'\eta}=-\frac{\sqrt{2}}{3}\,g_{f_0K^+K^-}\,x_{f_0}\,;$$

$$g_{\sigma\eta \eta}=g_{\sigma\eta \eta '}=\frac{\sqrt{2}}{3}\,
g_{\sigma \pi^+\pi^-}x_\sigma,\,\,g_{\sigma\eta '\eta
'}=\frac{1}{3\sqrt{2}}\,g_{\sigma \pi^+\pi^-}x_\sigma\,.$$

In the $K^+K^-$ loop model $g(m)$ has the following forms (see
Refs. \cite{achasov-89,achasov-01a,achasov-95,our_a0}).

\ \ \  For $m<2m_{K^+}$
\begin{eqnarray}
&&g(m)=\frac{e}{2(2\pi)^2}g_{\phi K^+K^-}\Biggl\{
1+\frac{1-\rho^2(m^2)}{\rho^2(m^2_{\phi})-\rho^2(m^2)}\times\nonumber\\
&&\Biggl[2|\rho(m^2)|\arctan\frac{1}{|\rho(m^2)|}
-\rho(m^2_{\phi})\lambda(m^2_{\phi})+i\pi\rho(m^2_{\phi})-\nonumber\\
&&-(1-\rho^2(m^2_{\phi}))\Biggl(\frac{1}{4}(\pi+
i\lambda(m^2_{\phi}))^2- \nonumber\\
&&-\Biggl(\arctan\frac{1}{|\rho(m^2)|}\Biggr)^2
\Biggr)\Biggr]\Biggr\},
\end{eqnarray}
where
\begin{equation}
\rho(m^2)=\sqrt{1-\frac{4m_{K^+}^2}{m^2}}\,\,;\qquad
\lambda(m^2)=\ln\frac{1+\rho(m^2)}{1-\rho(m^2)}\,\,;\qquad
\frac{e^2}{4\pi}=\alpha=\frac{1}{137}\,\,.
\end{equation}

 For $m\geq 2m_{K^+}$
\begin{eqnarray}
&&g(m)=\frac{e}{2(2\pi)^2}g_{\phi K^+K^-}\Biggl\{
1+\frac{1-\rho^2(m^2)}{\rho^2(m^2_{\phi})-\rho^2(m^2)}\times\nonumber\\
&&\times\Biggl[\rho(m^2)(\lambda(m^2)-i\pi)-
\rho(m^2_{\phi})(\lambda(m^2_{\phi})-i\pi)-\nonumber\\
&&\frac{1}{4}(1-\rho^2(m^2_{\phi}))
\Biggl((\pi+i\lambda(m^2_{\phi}))^2-
(\pi+i\lambda(m^2))^2\Biggr)\Biggr]\Biggr\}.
\end{eqnarray}

The mass spectrum of the reaction is
\begin{equation}
\frac{\Gamma(\phi\to\pi^0\pi^0\gamma)}{dm}=\frac{d\Gamma_{S}}{dm}+
\frac{d\Gamma_{back}(m)}{dm}+\frac{d\Gamma_{int}(m)}{dm},
\end{equation}
where the signal contribution $\phi\to S\gamma \to\pi^0\pi^0\gamma
$
 \begin{equation}
\frac{d\Gamma_{S}}{dm}=\frac{|P_K|^2
|g(m)|^2\sqrt{m^2-4m_{\pi}^2}(m_{\phi}^2-m^2)}
{3(4\pi)^3m_{\phi}^3}|\sum_{R,R'}g_{RK^+K^-}G_{RR'}^{-1}g_{R'\pi^0\pi^0}|^2.
\label{f0}
\end{equation}

The mass spectrum of the background process
$\phi\to\rho\pi^0\to\pi^0\pi^0\gamma$

\begin{equation}
\frac{d\Gamma_{back}(m)}{dm}=\frac{1}{2}\frac{(m_{\phi}^2-m^2)\sqrt{m^2-4m_{\pi}^2}}
{256\pi^3m_{\phi}^3} \int_{-1}^{1}dxA_{back}(m,x)\,,
\label{phonf0}
\end{equation}

where

\begin{eqnarray}
&&A_{back}(m,x)=\frac{1}{3}\sum|M_{back}|^2= \label{aback}\\
&&=\frac{F_b^2}{24}g_{\phi\rho\pi}^2g_{\rho\pi\gamma}^2 \bigg\{
\Big(m_{\pi}^8+2m^2m_{\pi}^4\tilde{m}_{\rho}^2-
4m_{\pi}^6\tilde{m}_{\rho}^2+2m^4\tilde{m}_{\rho}^4-\nonumber \\
&&-4m^2m_{\pi}^2\tilde{m}_{\rho}^4+
6m_{\pi}^4\tilde{m}_{\rho}^4+2m^2\tilde{m}_{\rho}^6-4m_{\pi}^2\tilde{m}_{\rho}^6+
\tilde{m}_{\rho}^8-2m_{\pi}^6m_{\phi}^2-\nonumber \\
&&-2m^2m_{\pi}^2\tilde{m}_{\rho}^2 m_{\phi}^2+
2m_{\pi}^4\tilde{m}_{\rho}^2m_{\phi}^2-2m^2\tilde{m}_{\rho}^4m_{\phi}^2+
2m_{\pi}^2\tilde{m}_{\rho}^4m_{\phi}^2-2\tilde{m}_{\rho}^6m_{\phi}^2+\nonumber
\\ &&+m_{\pi}^4m_{\phi}^4+ \tilde{m}_{\rho}^4m_{\phi}^4\Big)
\bigg(\frac{1}{|D_{\rho}(\tilde{m}_{\rho})|^2}+
\frac{1}{|D_{\rho}(\tilde{m}_{\rho}^*)|^2}\bigg)+\big(m_{\phi}^2-m^2\big)\big(m^2-\nonumber\\
&& -2m_{\pi}^2+2\tilde{m}_{\rho}^2-m_{\phi}^2\big)
\big(2m^2m_{\pi}^2+2m_{\pi}^2m_{\phi}^2-m^4\big)\frac{1}{|D_{\rho}(\tilde{m}_{\rho}^*)|^2}+
\nonumber \\
&&+2\mbox{Re}\bigg(\frac{1}{D_{\rho}(\tilde{m}_{\rho})D^*_{\rho}(\tilde{m}_{\rho}^*)}\bigg)
\Big(m_{\pi}^8-
m^6\tilde{m}_{\rho}^2+2m^4m_{\pi}^2\tilde{m}_{\rho}^2+ \nonumber
\\ &&+2m^2m_{\pi}^4\tilde{m}_{\rho}^2-4m_{\pi}^6\tilde{m}_{\rho}^2-
4m^2m_{\pi}^2\tilde{m}_{\rho}^4+6m_{\pi}^4\tilde{m}_{\rho}^4+\nonumber
\\ &&+2m^2\tilde{m}_{\rho}^6-4m_{\pi}^2\tilde{m}_{\rho}^6+\tilde{m}_{\rho}^8+
m^2m_{\pi}^4m_{\phi}^2-2m_{\pi}^6m_{\phi}^2+
2m^4\tilde{m}_{\rho}^2 m_{\phi}^2-\nonumber \\
&&-4m^2m_{\pi}^2\tilde{m}_{\rho}^2m_{\phi}^2+
2m_{\pi}^4\tilde{m}_{\rho}^2m_{\phi}^2-m^2\tilde{m}_{\rho}^4m_{\phi}^2+
2m_{\pi}^2\tilde{m}_{\rho}^4m_{\phi}^2-2\tilde{m}_{\rho}^6m_{\phi}^2-\nonumber
\\ &&-m_{\pi}^4m_{\phi}^4-m^2\tilde{m}_{\rho}^2m_{\phi}^4+
2m_{\pi}^2\tilde{m}_{\rho}^2m_{\phi}^4+\tilde{m}_{\rho}^4m_{\phi}^4\Big)
\bigg\},\nonumber
\end{eqnarray}

\begin{eqnarray}
&&\tilde{m_{\rho}}^2=m_{\pi}^2+\frac{(m_{\phi}^2-m^2)}{2}
(1-x\sqrt{1-\frac{4m_{\pi}^2}{m^2}})\nonumber \\
&&\tilde{m_{\rho}}^{*2}=m^2_{\phi}+2m_{\pi}^2-m^2-\tilde{m_{\rho}}^2\,.
\end{eqnarray}

The interference between signal and background processes accounts
for

\begin{equation}
\frac{d\Gamma_{int}(m)}{dm}=\frac{1}{\sqrt{2}}\frac{\sqrt{m^2-4m_{\pi}^2}}
{256\pi^3m_{\phi}^3} \int_{-1}^{1}dxA_{int}(m,x)\,,
 \label{intf0}
\end{equation}
where

\begin{eqnarray}
&&A_{int}(m,x)=\frac{2}{3}(m_\phi^2-m^2)\mbox{Re}\sum M_f
M_{back}^*= \label{f0int}\\ &&=\frac{16\pi}{3}F_b
\mbox{Re}\Biggl\{e^{i\delta}g(m)g_{\phi\rho\pi}g_{\rho\pi\gamma}
T^0_0\left (K^+K^-\to\pi^0\pi^0\right ) \Biggl [
\frac{(\tilde{m}_{\rho}^2-m_{\pi}^2)^2m_{\phi}^2-(m_{\phi}^2-m^2)^2\tilde{m}_{\rho}^2
} {D_{\rho}^*(\tilde{m}_{\rho})}+\nonumber\\
&&+\,\frac{(\tilde{m}_{\rho}^{*2}-m_{\pi}^2)^2m_{\phi}^2-(m_{\phi}^2-m^2)^2
\tilde{m}_{\rho}^{*2} }{D_{\rho}^*(\tilde{m}_{\rho}^*)}\Biggr ]
\Biggr\}=\nonumber\\
 &&=\frac{F_b}{3}\mbox{Re}\Biggl\{P_K
e^{i\delta_B^{\pi\pi}}e^{i\delta}g(m)g_{\phi\rho\pi}g_{\rho\pi^0\gamma}
\Biggl (\sum_{R,R'}g_{RK^+K^-}G_{RR'}^{-1}g_{R'\pi^0\pi^0}\Biggr
)\times\nonumber\\ &&\times\Biggl
[\frac{(\tilde{m}_{\rho}^2-m_{\pi}^2)^2m_{\phi}^2-(m_{\phi}^2-m^2)^2\tilde{m}_{\rho}^2
} {D_{\rho}^*(\tilde{m}_{\rho})}
+\frac{(\tilde{m}_{\rho}^{*2}-m_{\pi}^2)^2m_{\phi}^2-(m_{\phi}^2-m^2)^2
\tilde{m}_{\rho}^{*2}}{D_{\rho}^*(\tilde{m}_{\rho}^*)}\Biggr ]
\Biggr\}\,. \nonumber
\end{eqnarray}

The factor $1/2$ in Eq. (\ref{phonf0}) and the factor $1/\sqrt{2}$
in Eq. (\ref{intf0}) take into account the identity of pions.

The S-wave amplitude $T^0_0$ of the $\pi\pi$ scattering with I=0
\cite{adsh-1980,z_phys,adsh-84,achasov-97} is

\begin{equation}
T^0_0=\frac{\eta^0_0 e^{2i\delta_0^0}-1}{2i\rho_{\pi\pi}(m)}=
\frac{e^{2i\delta_B^{\pi\pi}}-1}{2i\rho_{\pi\pi}(m)}+
e^{2i\delta_B^{\pi\pi}}\sum_{R,R'}\frac{g_{R\pi\pi}G_{RR'}^{-1}g_{R'\pi\pi}}{16\pi}\,.
\label{pipiamp}\end{equation}

Here $\eta^0_0\equiv \eta^0_0(m)$ is the inelasticity,
$\eta^0_0=1$ for $m\leq 2m_{K^+}$, and

\begin{equation}
\label{phas2}
\delta_0^0\equiv\delta_0^0(m)=\delta^{\pi\pi}_B(m)+\delta_{res}(m)\,,
\end{equation}

\noindent where $\delta_B^{\pi\pi}=\delta_B^{\pi\pi}(m)$
($\delta_B$ in Ref. \cite{achasov-97}) is the phase of the elastic
background [see Eq. \ref{delph}], and $\delta_{res}(m)$ is the
resonance scattering phase,

\begin{equation}
S_0^{0\ res}=\eta^0_0(m) e^{2i\delta_{res}(m)}=1+2i
\rho_{\pi\pi}(m)
\sum_{R,R'}\frac{g_{R\pi\pi}G_{RR'}^{-1}g_{R'\pi\pi}}{16\pi}\,,\,\,\,\eta^0_0=|S_0^{0\
res}|\,, \label{phR} \end{equation}

\noindent $g_{R\pi\pi}=\sqrt{3/2}\,g_{R\pi^+\pi^-}$. The chiral
shielding phase $\delta_B^{\pi\pi}(m)$, motivated by the $\sigma$
model \cite{annshgn-94,annshgn-07} and desired analytical
properties, is taken in more complicated form than in Ref.
\cite{our_f0}, see Sec. \ref{sBackPhase}.

The phase $\delta_B^{K \bar K}=\delta_B^{K \bar K}(m)$ is
parametrized in the following way:

\begin{equation}
\tan \delta_B^{K \bar K}=f_K(m^2)\sqrt{m^2-4m^2_{K^+}}\equiv 2p_K
f_K(m^2)\label{phK}
\end{equation}
\noindent and

\begin{equation}
e^{2i\delta_B^{K\bar K}}=\frac{1+i2p_K f_K(m^2)}{1-i2p_K
f_K(m^2)}\label{phK2}
\end{equation}

Actually, $e^{2i\delta^{\pi\pi}_B(m)}$ has a pole at $m^2=m_0^2$,
$ 0<m^2_0<4m^2_\pi$, which is compensated by the zero in $e^{2i
\delta_B^{K \bar K}(m)}$ to ensure a regular $K\bar K\to\pi\pi$
amplitude and, consequently, the $\phi\to K^+K^-\to\pi\pi\gamma$
amplitude at $ 0<m^2<4m_\pi^2$. This requirement leads to
\begin{equation}
f_K(m_0^2)=\frac{1}{\sqrt{4m^2_{K^+}-m_0^2}}\approx
\frac{1}{2m_{K^+}}\,. \label{kcondition}
\end{equation}

As in Refs. \cite{our_f0}, for $f_K(m^2)$ we used the form

\begin{equation}
f_K(m^2)=-\frac{\arctan (\frac{m^2-m_1^2}{m_2^2})}{\Lambda _K}\,.
\label{fK}
\end{equation}

The inverse propagator of the $\rho$ meson has the following
expression:
\begin{equation}
D_{\rho}(m)=m_{\rho}^2-m^2-im^2\frac{g^2_{\rho\pi\pi}}{48\pi}
\bigg(1-\frac{4m_{\pi}^2}{m^2}\bigg)^{3/2}\,.
\end{equation}.

The coupling constants $g_{\phi K^+K^-}=4.376\pm 0.074$ and
$g_{\phi\rho\pi}=0.814\pm 0.018$ GeV$^{-1}$ are taken from the
most precise measurement \cite{sndphi}. To obtain the coupling
constant $g_{\rho\pi^0\gamma}$ we used the data of the experiments
\cite{dolinsky} and \cite{pigam} on the $\rho\to\pi^0\gamma$ decay
and the expression

\begin{equation}
\Gamma(\rho\to\pi^0\gamma)=\frac{g_{\rho\pi^0\gamma}^2}{96\pi
m_{\rho}^3} (m_{\rho}^2-m_{\pi}^2)^3,
\end{equation}

\noindent the result $g_{\rho\pi^0\gamma}=0.26\pm 0.02$ GeV$^{-1}$
is the weighed average of these experiments.

\section{The background phase $\delta_B^{\pi\pi}$ }
\label{sBackPhase}

The proper analytical properties of the $\pi\pi$ scattering
amplitude are two cuts in the $s$-complex plane, Adler zero in
$T^0_0$ \cite{AdlerZero}, absence of poles on the physical sheet
of the Riemannian surface, $\sigma(600)$ and $f_0(980)$ poles in
the resonance amplitude on the second sheet of the Riemannian
surface, and absence of poles on the second sheet in the
background amplitude in the region $4m_\pi^2<\mbox{Re} (s) < (1.2$
GeV$)^2$. This applies curtain restrictions on the
$\delta_B^{\pi\pi}$.

Let us represent $\delta_B^{\pi\pi}$ in the physical region $s =
m^2> 4m_\pi^2$ as

\begin{equation}\tan ( \delta_B^{\pi\pi})=\frac{\mbox{Im}\,{(P_{\pi 1}(s)P_{\pi 2}(s))}}{\mbox{Re}\,{(P_{\pi 1}(s)P_{\pi 2}(s))}}\,,\label{phB}\end{equation}

\noindent and
\begin{equation}
e^{2i\delta_B^{\pi\pi}}=S_1^{back}S_2^{back}=\frac{P^*_{\pi
1}(s)P^*_{\pi 2}(s)}{P_{\pi 1}(s)P_{\pi 2}(s)}=\frac{P_{\pi
1}(s-i\epsilon)P_{\pi 2}(s-i\epsilon)}{P_{\pi
1}(s+i\epsilon)P_{\pi 2}(s+i\epsilon)}\,, \label{phB2}
\end{equation}

\noindent where

\begin{equation}
P_{\pi 1}(s) =
a_1-a_2\frac{s}{4m_\pi^2}-\Pi_{\pi\pi}(s)+a_3\,\Pi_{\pi\pi}(4m_\pi^2-s)-a_4
Q_1(s)\,,
\end{equation}

\begin{equation}
Q_1(s)
=\frac{1}{\pi}\int_{4m_\pi^2}^{\infty}\frac{s-4m_\pi^2}{s'-4m_\pi^2}\frac{\rho_{\pi\pi}(s')}{s'-s-i\varepsilon}K_1(s')\,,
\end{equation}

\begin{equation}
K_1(s) =
\frac{L_1(s)}{D_1(4m_\pi^2-s)D_2(4m_\pi^2-s)D_3(4m_\pi^2-s)D_4(4m_\pi^2-s)D_5(4m_\pi^2-s)D_6(4m_\pi^2-s)}\,,
\label{kernel1}
\end{equation}

$$L_1(s) = (s-4m_\pi^2)^6+\alpha_1 (s-4m_\pi^2)^5+\alpha_2
(s-4m_\pi^2)^4 +\alpha_3 (s-4m_\pi^2)^3+$$ $$+\alpha_4
(s-4m_\pi^2)^2+\alpha_5 (s-4m_\pi^2)+\alpha_6+$$ $$
+\sqrt{s}\bigg(c_1(s-4m_\pi^2)^5 +
c_2(s-4m_\pi^2)^4+c_3(s-4m_\pi^2)^3+$$
\begin{equation}
+c_4(s-4m_\pi^2)^2+c_5(s-4m_\pi^2)+c_6\bigg)\,,
\end{equation}

\begin{equation}
D_i(s)=m_{i}^2-s-g_i\Pi_{\pi\pi}(s)\,,
\end{equation}

\begin{equation}
\Pi_{\pi\pi}(s)=\frac{16\pi}{g_{Rab}^2}\Pi_R^{\pi\pi}(s)\,,
\end{equation}

\begin{equation}
P^*_{\pi 1}(s) = P_{\pi 1}(s-i\epsilon)=P_{\pi
1}(s)+2i\rho_{\pi\pi} (s)\bigg(1+a_4K_1(s)\bigg)\,,
\end{equation}

\begin{equation}
P_{\pi 2}(s) = \frac{\Lambda^2+s-4m_\pi^2}{4m_\pi^2} + k_2Q_2(s)\,,
\end{equation}

\noindent here

\begin{equation}
Q_2(s)
=\frac{1}{\pi}\int_{4m_\pi^2}^{\infty}\frac{s-4m_\pi^2}{s'-4m_\pi^2}\frac{\rho_{\pi\pi}(s')}{s'-s-i\varepsilon}K_2(s')\,,
\end{equation}

\begin{equation}
K_2(s) =
\frac{L_2(s)}{D_{1A}(4m_\pi^2-s)D_{2A}(4m_\pi^2-s)D_{3A}(4m_\pi^2-s)}\,,
\label{kernel2}
\end{equation}

\begin{equation}
L_2(s) = 4m_\pi^2\bigg(s^2+\beta s + \gamma_1 s^{3/2} + \gamma_2
s^{1/2}\bigg)\,,
\end{equation}

\begin{equation}
P^*_{\pi 2}(s) = P_{\pi 2}(s-i\epsilon)= P_{\pi 2}(s) -
2i\rho_{\pi\pi} (s)k_2 K_2(s)\,.
\end{equation}

Note that this parametrization was inspired by Ref.
\cite{ourProp}, devoted to proof that the propagators
(\ref{propagator}) satisfy the K\"allen-Lehmann representation in
the wide domain of coupling constants of the scalar mesons with
the two-particle states. Following the ideas of this paper the
conditions

$$K_1(s)\geq 0,\, K_2(s)\geq 0\ \mbox{at}\ s>4m_\pi^2$$

\noindent guarantee absence of poles on the physical sheet in Eq.
(\ref{phB2}) (of course, the restrictions of Sec. \ref{restr}
should be fulfilled too). Note also that we choose the denominator
of (\ref{phB2}) as $P_{\pi 1}(s) P_{\pi 2}(s)$ for our comfort.


\section{Restrictions on the parameters} \label{restr}

Some parameters are fixed by the requirement of the proper
analytical continuation of amplitudes. The denominators $P_{\pi
1}$ and $P_{\pi 2}$ have zeroes at $s=m_0^2$ and $s=m_{0A}^2$
respectively, both belonging to the interval $0<s<4m^2_\pi$. These
zeroes should be compensated by zeroes in any pair from $P^*_{\pi
1}$, $P^*_{\pi 2}$ and $S_0^{0\ res}$. We choose

$$P^*_{\pi 1}(m_{0}^2)=0\,,$$

\begin{equation}
S_0^{0\ res}(m_{0A}^2)=0\,, \label{cond2}
\end{equation}

\noindent see Eq. (\ref{phB2}) \cite{SimilarConditions}.

The requirement of the $T^0_0$ finiteness at $s=0$ leads to $2$
conditions. Really, on the real axis for $s>4m_\pi^2$ we have

$$S_1^{back}=\frac{P^*_{\pi 1}(s)}{P_{\pi 1}(s)}=\frac{P_{\pi
1}(s-i\epsilon)}{P_{\pi
1}(s+i\epsilon)}=1+2i\rho_{\pi\pi}(s)\frac{1+K_1(s)}{P_{\pi
1}(s)}\,,$$

$$S_2^{back}=\frac{P^*_{\pi 2}(s)}{P_{\pi 2}(s)}=\frac{P_{\pi
2}(s-i\epsilon)}{P_{\pi
2}(s+i\epsilon)}=1-2i\rho_{\pi\pi}(s)\frac{K_2(s)}{P_{\pi
2}(s)}\,.$$

So, to avoid singularity in the

$$T_0^0=\frac{S_1^{back}S_2^{back}S^{0\,res}_0-1}{2i\rho_{\pi\pi}(s)}$$

\noindent at $s=0$, where $\rho_{\pi\pi}(s)$ becomes infinite, we
require

$$1+K_1(0)=0,$$

\noindent as for $K_2(0)$, it is equal to zero at $s=0$ via
construction, see Eq. (\ref{kernel2}). Note that, alternatively,
one may require $T^{0\,res}_0(0)=0$.

Additionally, as it was noted in Refs. \cite{nullDerivIn0},
crossing symmetry implemented by Roy equations imposes the
condition

$$\frac{dT^0_0}{dm} (m^2=0) = 0\,.$$

Recall that the condition Eq. (\ref{kcondition}) removes the
singularity in the $T(\pi\pi\to K\bar K)$ amplitude. One can see
that no special prerequisite to Adler zero existence in the
$\pi\pi$ scattering amplitude should be imposed, because it
appears when we take into account the results of Ref.
\cite{sigmaPole}.

\section{Data analysis}
\label{sr}

Analyzing data, we imply a scenario motivated by the four-quark
model \cite{jaffe}, that is, the $\sigma$(600) coupling with the
$K\bar K$ channel, $g_{\sigma K^+K^-}$, is suppressed relatively
to the coupling with the $\pi\pi$ channel, $g_{\sigma
\pi^+\pi^-}$, the mass of the $\sigma$ meson $m_\sigma$ is in the
500-700 MeV range. In addition, we have in mind the Adler
self-consistency conditions for the $T_0^0(\pi\pi\to\pi\pi)$ near
the $\pi\pi$ threshold. The general aim of this section is to
demonstrate that the data and the \cite{sigmaPole} results on the
$\pi\pi$ amplitude are in excellent agreement with this general
scenario.

As in Ref. \cite{our_f0} for $\phi\to\pi^0\pi^0\gamma$ decay we
use the KLOE data \cite{pi0publ} for $m>660$ MeV. For the
$\delta_0^0$ we use the "old data"
\cite{hyams,estabrook,martin,srinivasan,rosselet}, 44 points up to
$1.2$ GeV \cite{theSameData}. Besides, we take into account the
new precise data in the low energy region \cite{scatbnl,na48}.

The inelasticity $\eta^0_0(m)$ and the phase $\delta ^{\pi K}(m)$
of the amplitude $T(\pi\pi\to K\bar K)$ are essential in the fit
region, $2m_{K^+} < m < 1.2$ GeV. As for the inelasticity, the
experimental data of Ref. \cite{hyams} gives evidence in favor of
low values of $\eta^0_0(m)$ near the $K\bar K$ threshold. At
present the contribution of the $\eta\eta,\,\eta'\eta$, and
$\eta'\eta'$ channels does not affect much the overall fits. To
fix a relation between the $K\bar K$ and $\eta\eta$ channels
reliably the inelasticity should be measured with accuracy 10
times better than the existing one. The situation with the
experimental data on $\delta ^{\pi K}(m)$ is controversial and
experiments have large errors. We consider these data as a guide,
whose main role is to fix the sign between signal (\ref{f0signal})
and background amplitudes (\ref{amp_back}), and hold two points of
the experiment \cite{pkphase}, see Fig. \ref{fig10}. As for
inelasticity, for fitting we used only the key experimental point
$\eta^0_0 (m=1.01 \mbox{ GeV} )= 0.41\pm 0.14$, see Fig.
\ref{fig6}.

Providing all the above conditions, we have obtained perfect
agreement with the general scenario under consideration, see Fits
1, 2 in Tables I, II, and III and Figs. 1-10. Fits 1 and 2 show
that the allowed range of $\sigma(600)$ and $f_0(980)$ parameters
is rather wide.

The values of $g_{f_0K^+K^-}^2/4\pi$ in Fits 1 and 2 ($1$ GeV$^2$
and $2$ GeV$^2$, correspondingly) show a scale of possible
deviation of this constant. This may be important to coordinate
$g_{f_0K^+K^-}^2/4\pi$ with $g_{a_0K^+K^-}^2/4\pi$
\cite{equalConstants}, note the latter is usually larger than 1
GeV$^2$.

In addition, we carry out Fit 3, where $\sigma(600)$ and
$f_0(980)$ are coupled only with the $\pi\pi$ channel. As seen
from Table I and Figs. 11-13, Fit 3 is in excellent agreement with
the data on the $\delta_0^0$ up to 1 GeV and the \cite{sigmaPole}
results.

We introduce 52 parameters, but for restrictions (expresses 5
parameters through others) and parameters (or their combinations),
that go to the bound of the permitted range (7 effective links),
the effective number of free parameters is reduced to 40. It is
significant that fits describe not only the experimental data
(about 80 points), but also the $\pi\pi$ amplitude from the
\cite{sigmaPole} in the range $-5m_\pi^2 < s < 0.64$ GeV$^2$ which
is treated along with experimental data.

The $\sigma (600)$ pole positions, obtained in Fits 1 and 2, lie
far from Eq. (\ref{poleSigma}), see Table I. One of the possible
reasons is neglecting $K\bar K$ and other high channels in the
\cite{sigmaPole} approach. The role of high channels can be
estimated with the help of Fit 3, whose $\sigma(600)$ pole
position is considerably closer to Eq. (\ref{poleSigma}), see
Table I.

Note that kernels of the background integrals (\ref{kernel1}) and
(\ref{kernel2}) are positive in the range of integration [$2m_\pi,
\infty)$, Fit 1 kernels are presented in Fig. 7.

The Adler zero in the $T^0_0(\pi\pi\to \pi\pi)$ is near $s=(100$
MeV$)^2$ in all Fits because we describe the amplitude
\cite{sigmaPole}. Fit 2 also has Adler zero in the $T(\pi\pi\to
K\bar K)$ at $s=(166$ MeV$)^2$, Fit 1 has a zero in the
$T(\pi\pi\to K\bar K)$ at $s=-(601$ MeV$)^2$.

The resonance amplitude $T^{0\,res}_0$ has poles on the unphysical
sheets of its Riemannian surface. As we have a multichannel case,
the amplitude has the set of lists depending on lists of the
polarization operators $\Pi_R^{ab}(s)$. We show resonance poles
only on some lists, see Tables IV and V. For this choice, in case
of metastable states, decaying to several channels, the imaginary
parts of pole positions $M_R$ would be connected to the full
widths of the resonances [$2\mbox{Im}M_R=\Gamma_R=\sum_{ab}\Gamma
(R\to ab)$]. Note that $\sigma(600)$ and $f_0(980)$ poles, shown
in Table I, correspond to the first lines of Tables IV and V.

As to the background amplitude $T^{0\,back}_0$, it has poles on
the second sheet of the Riemannian surface, where $P_{\pi 1} = 0$
or $P_{\pi 2} = 0$. The $P_{\pi 1}$ has a zero at
$s=(1246-104\,i)^2$ MeV$^2$ for Fit 1, at $s=(1354-110\,i)^2$
MeV$^2$ for Fit 2, and at $s=(1056-142\,i)^2$ MeV$^2$ for Fit 3.
The $P_{\pi 2}$ has a zero at $s=(0.2 - 9.5\,i)\,m^2_\pi$ for Fit
1, at $s=(2.0-8.9\,i)\,m^2_\pi$ for Fit 2, and at
$s=(-0.6-8.6\,i)\,m^2_\pi$ for Fit 3. These poles lie outside of
the region $4m_\pi^2<\mbox{Re}(s)<(1.2$ GeV$)^2$ except the pole
at $s=(1056-142\,i)^2$ MeV$^2$ for Fit 3, but for this Fit the
upper bound is $1$ GeV$^2$.\\[-1pt]

\begin{center}
\vspace{-15pt}Table I. Properties of the resonances and main
characteristics are shown. \vspace{2pt}
\begin{tabular}{|c|c|c|c|}\hline

Fit & 1 & 2 & 3 \\ \hline $m_{f_0}$, MeV & $979.16$ & $986.50$ &
$964.01$  \\ \hline

$g_{f_0K^+K^-}$, GeV  & $3.54$ & $5.01$ & $0$   \\ \hline

$\frac{g_{f_0K^+K^-}^2}{4\pi}$, GeV$^2$  & $1$ & $2$ & $0$ \\
\hline

$g_{f_0 \pi^+\pi^-}$, GeV  & $-1.3737$ & $-2.1185$ & $0.3183$
\\ \hline

$\frac{g_{f_0\pi^+ \pi^-}^2}{4\pi}$, GeV$^2$ & $0.150$ & $0.357$ &
$0.008$   \\ \hline

$x_{f_0}$ & $0.6640$ & $0.9584$ & -   \\ \hline

$\Gamma_{f_0}(m_{f_0})$, MeV & $55.2$ & $130.3$ & $3.0$
\\ \hline

$f_0(980)$ pole, MeV & $986.2 - 25.5\,i$ & $990.5 - 19.4\,i$ &
$978.9 - 11.4\,i$
\\ \hline

$m_{\sigma}$, MeV & $487.59$ & $506.95$ & $480.46$   \\ \hline

$g_{\sigma\pi^+ \pi^-}$, GeV & $2.7368$ & $2.6735$ & $2.5871$
\\ \hline

$\frac{g_{\sigma\pi^+ \pi^-}^2}{4\pi}$, GeV$^2$ & $0.596$ &
$0.569$ & $0.533$   \\ \hline

$g_{\sigma K^+K^-}$, GeV & $0.552$ & $0.774$ & $0$
\\ \hline

$\frac{g_{\sigma K^+K^-}^2}{4\pi}$, GeV$^2$ & $0.024$ & $0.048$ &
$0$   \\ \hline

$x_\sigma$ & $0.9750$ & $0.8201$ & $0$   \\ \hline

$\Gamma_{\sigma}(m_{\sigma})$, MeV & $377.8$ & $352.9$ & $340.2$
\\ \hline

$\sigma(600)$ pole, MeV & $581.0 - 212.7\,i$ & $613.8 - 221.4\,i$
& $528.6 - 220.3\,i$
\\ \hline

$C$, GeV$^2$ & $0.04317$ & $-0.07633$ & $-0.11734$  \\ \hline

$\delta $, $^{\circ}$ & $-70.62$& $-73.6$ & -   \\ \hline

$m_1$, MeV & $801.90$ & $814.88$ & -  \\ \hline

$m_2$, MeV & $465.95$ & $554.95$ & -  \\ \hline

$\Lambda _K$, GeV & $1.142$ & $1.030$ & - \\ \hline

$a^0_0,\ m_\pi^{-1}$ & $0.223$ & $0.226$ & $0.221$ \\ \hline

Adler zero in $\pi\pi\to\pi\pi $ & ($94.4$ MeV)$^2$ & ($96.8$
MeV)$^2$ & ($87.1$ MeV)$^2$  \\ \hline

$\eta^0_0$($1010$ MeV) & $0.55$ & $0.45$ & -  \\ \hline

$\chi^2_{phase}$ (44 points) & $45.9$ & $50.6$ & $26.3$ ($34$
points)
\\ \hline

$\chi^2_{sp}$ ($18$ points) & $24.9$ & $19.1$ & -
\\ \hline
\end{tabular}
\end{center}

\newpage

\begin{center}
\vspace{-15pt}Table II. Parameters of the first background
($P_{\pi 1}$) are shown. \vspace{2pt}
\begin{tabular}{|c|c|c|c|}\hline

Fit & 1 & 2 & 3 \\ \hline $a_1$ & $-3.105$ & $-4.549$ & $-1.498$
\\ \hline

$a_2$ & $0.01136$ & $0.00998$ & $0.05821$   \\ \hline

$a_3$ & $0$ & $0$ & $0$ \\ \hline

$a_4$  & $4.9328$ & $13.1111$ & $1.2475$
\\ \hline

$\alpha_1$, GeV$^2$ & $604.137$ & $624.512$ & $-792.804$   \\
\hline

$\alpha_2$, GeV$^4$ & $920.111$ & $1000.739$ & $-384.477$
\\ \hline

$\alpha_3$, GeV$^6$ & $785.958$ & $781.770$ & $416.645$
\\ \hline

$\alpha_4$, GeV$^8$ & $223.623$ & $211.195$ & $198.772$
\\ \hline

$\alpha_5$, GeV$^{10}$ & $24.5339$ & $23.8517$ & $25.4265$
\\ \hline

$\alpha_6$, GeV$^{12}$ & $0.248657$ & $0.314094$ & $0.198560$
\\ \hline

$c_1$, GeV & $356.128$ & $224.404$ & $995.905$
\\ \hline

$c_2$, GeV$^3$ & $-2735.40$ & $-2600.82$ & $-1070.75$   \\ \hline

$c_3$, GeV$^5$ & $284.008$ & $445.192$ & $542.745$
\\ \hline

$c_4$, GeV$^7$ & $430.758$ & $461.717$ & $411.927$
\\ \hline

$c_5$, GeV$^9$ & $49.7913$ & $47.2357$ & $51.4206$  \\ \hline

$c_6$, GeV$^{11}$ & $-0.664290$& $-0.684002$ & $-0.635647$   \\
\hline

$m_1$, MeV & $1105.67$ & $1111.87$ & $1002.31$  \\ \hline

$g_1$, MeV & $347.70$ & $350.48$ & $306.18$  \\ \hline

$m_2$, MeV & $1061.53$ & $1141.92$ & $806.93$  \\ \hline

$g_2$, MeV & $344.12$ & $381.73$ & $350.51$  \\ \hline

$m_3$, MeV & $1061.85$ & $1169.51$ & $781.76$  \\ \hline

$g_3$, MeV & $311.56$ & $311.80$ & $322.57$  \\ \hline

$m_4$, MeV & $970.78$ & $1040.96$ & $970.78$  \\ \hline

$g_4$, MeV & $457.52$ & $455.56$ & $376.88$  \\ \hline

$m_5$, MeV & $1176.39$ & $1320.55$ & $1153.21$  \\ \hline

$g_5$, MeV & $544.43$ & $588.48$ & $500.59$  \\ \hline

$m_6$, MeV & $1521.20$ & $1621.10$ & $1808.74$  \\ \hline

$g_6$, MeV & $739.93$ & $750.75$ & $841.57$  \\ \hline

\end{tabular}
\end{center}

\newpage

\begin{center}
\vspace{-15pt}Table III. Parameters of the second background
($P_{\pi 2}$) are shown. \vspace{2pt}
\begin{tabular}{|c|c|c|c|}\hline

Fit & 1 & 2 & 3 \\ \hline $\Lambda$, MeV & $83.238$ & $74.477$ &
$70.268$ \\ \hline

$k_2$ & $0.0152934$ & $0.0168176$ & $0.0150655$ \\ \hline

$\beta$ & $239.184$ & $221.055$ & $263.511$ \\ \hline

$\gamma_1$ & $1006.367$ & $928.743$ & $878.056$ \\ \hline

$\gamma_2$ & $22.7004$ & $23.3341$ & $29.4097$ \\ \hline

$m_{1A}$, MeV & $491.92$ & $84.77$ & $687.43$  \\ \hline

$g_{1A}$, MeV & $469.29$ & $492.03$ & $364.68$  \\ \hline

$m_{2A}$, MeV & $531.81$ & $639.95$ & $528.40$  \\ \hline

$g_{2A}$, MeV & $452.20$ & $261.48$ & $378.65$  \\ \hline

$m_{3A}$, MeV & $670.64$ & $565.16$ & $608.72$  \\ \hline

$g_{3A}$, MeV & $299.23$ & $428.97$ & $370.98$  \\ \hline

\end{tabular}
\end{center}\vspace{10pt}

\noindent \hspace{60pt}Table IV. $\sigma(600)$ poles (MeV) on
different sheets of the complex $s$
\begin{center}
\vspace{-15pt} plane depending on lists of polarization operators
$\Pi^{ab}(s)$ are shown.\vspace{2pt}
\begin{tabular}{|c|c|c|c|c|c|}\hline

$\Pi^{K\bar K}$ list & $\Pi^{\eta\eta}$ list & $\Pi^{\eta\eta'}$
list & $\Pi^{\eta'\eta'}$ list & Fit 1 & Fit 2 \\ \hline

I & I & I & I & $581.0 - 212.7\,i$ & $613.8 - 221.4\,i$
\\ \hline

II & I & I & I & $617.5 - 353.0\,i$ & $609.8 - 291.6\,i$
\\ \hline

II & II & I & I & $554.3 - 375.3\,i$ & $559.4 - 346.6\,i$
\\ \hline

II & II & II & I & $579.0 - 475.2\,i$ & $569.7 - 410.7\,i$
\\ \hline

II & II & II & II & $625.7 - 474.9\,i$ & $581.6 - 411.0\,i$
\\ \hline

\end{tabular}
\end{center}\vspace{10pt}

\noindent \hspace{60pt}Table V. $f_0(980)$ poles (MeV) on
different sheets of the complex $s$
\begin{center}
\vspace{-15pt} plane depending on lists of polarization operators
$\Pi^{ab}(s)$ are shown. \vspace{2pt}
\begin{tabular}{|c|c|c|c|c|c|}\hline

$\Pi^{K\bar K}$ list & $\Pi^{\eta\eta}$ list & $\Pi^{\eta\eta'}$
list & $\Pi^{\eta'\eta'}$ list & Fit 1 & Fit 2 \\ \hline

I & I & I & I & $986.2 - 25.5\,i$ & $990.5 - 19.4\,i$
\\ \hline

II & I & I & I & $916.9 - 299.4\,i$ & $1183.2 - 518.6\,i$
\\ \hline

II & II & I & I & $966.8 - 450.5\,i$ & $1366.0 - 756.5\,i$
\\ \hline

II & II & II & I & $962.6 - 465.2\,i$ & $1390.7 - 813.0\,i$
\\ \hline

II & II & II & II & $962.5 - 608.0\,i$ & $1495.6 - 1057.7\,i$
\\ \hline

\end{tabular}
\end{center}

\begin{figure}[h]
\begin{center}
\begin{tabular}{ccc}
\includegraphics[width=8cm]{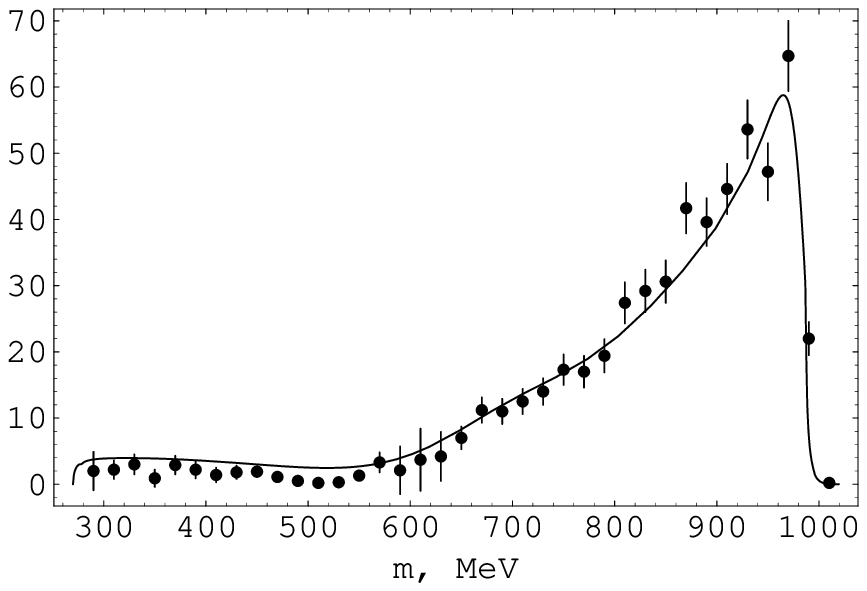}& \includegraphics[width=8cm]{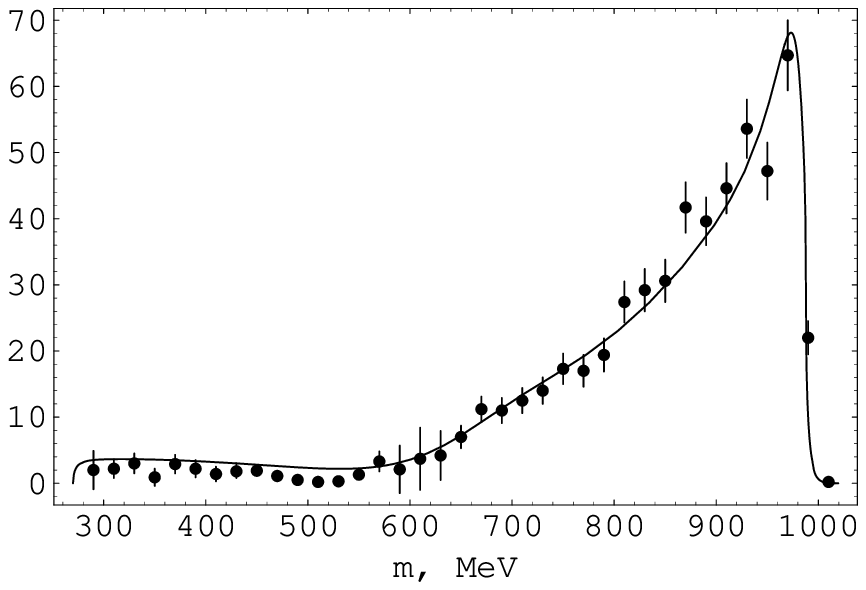}\\ (a)&(b)
\end{tabular}
\end{center}
\caption{The $\pi^0\pi^0$ spectrum in the
$\phi\to\pi^0\pi^0\gamma$ decay, theoretical curve, and the KLOE
data (points) are shown: a) Fit 1, b) Fit 2.} \label{fig2}
\end{figure}

\begin{figure}[h]
\begin{center}
\begin{tabular}{ccc}
\includegraphics[width=8cm]{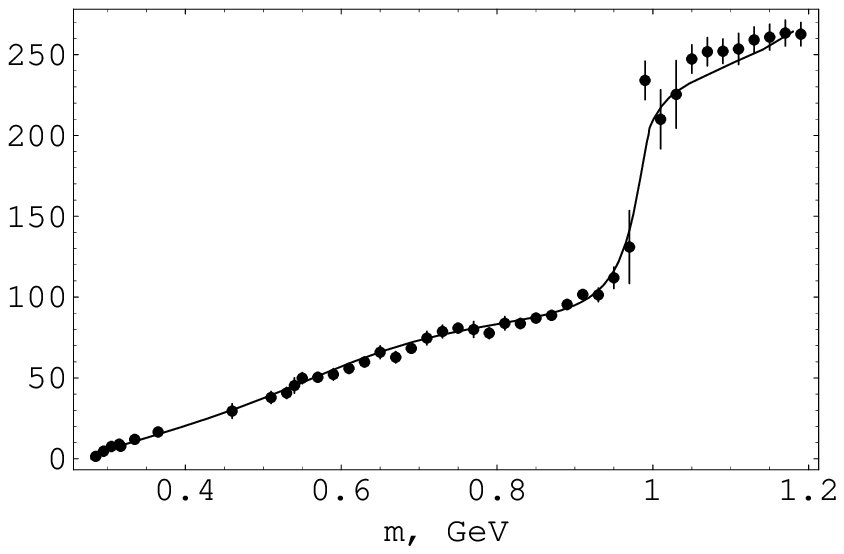}& \raisebox{-1mm}{$\includegraphics[width=8cm]{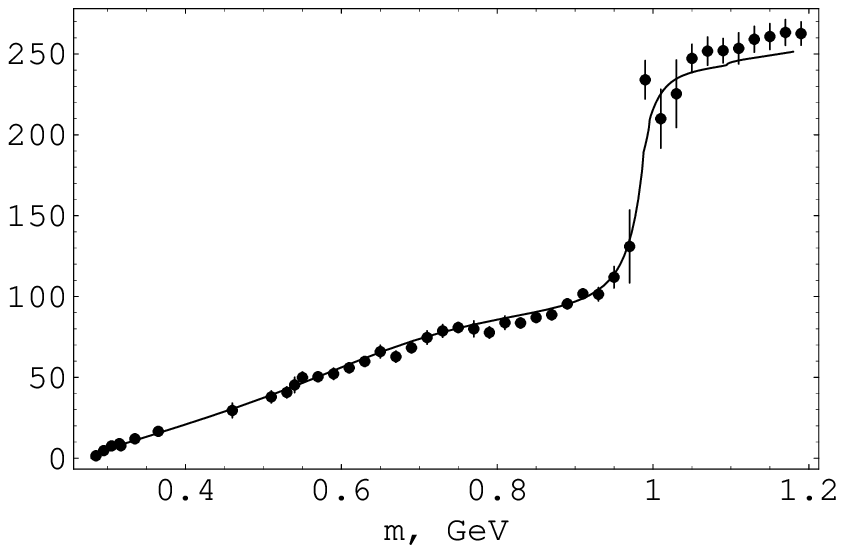}$}\\ (a)&(b)
\end{tabular}
\end{center}
\caption{The phase $\delta_0^0$ of the $\pi\pi$ scattering
(degrees) is shown: a) Fit 1, b) Fit 2.} \label{fig3}
\end{figure}

\begin{figure}[h]
\begin{center}
\begin{tabular}{ccc}
\includegraphics[width=8cm]{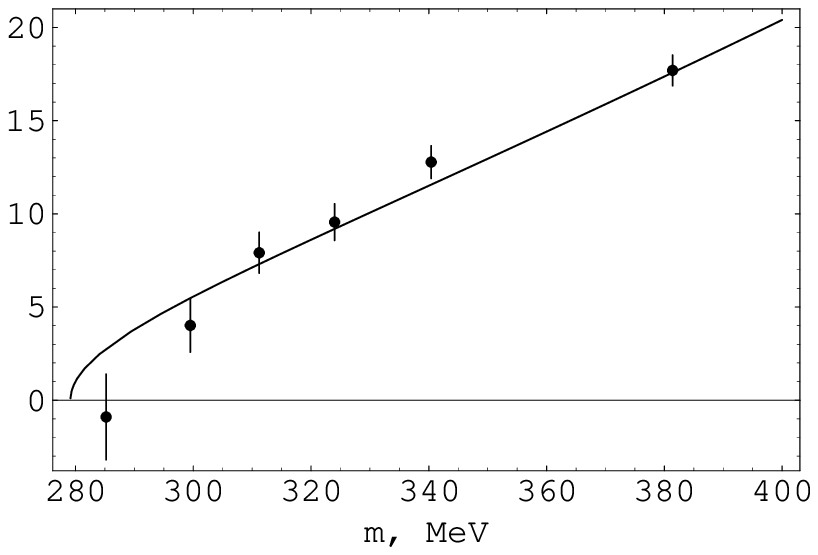}& \raisebox{-1mm}{$\includegraphics[width=8cm]{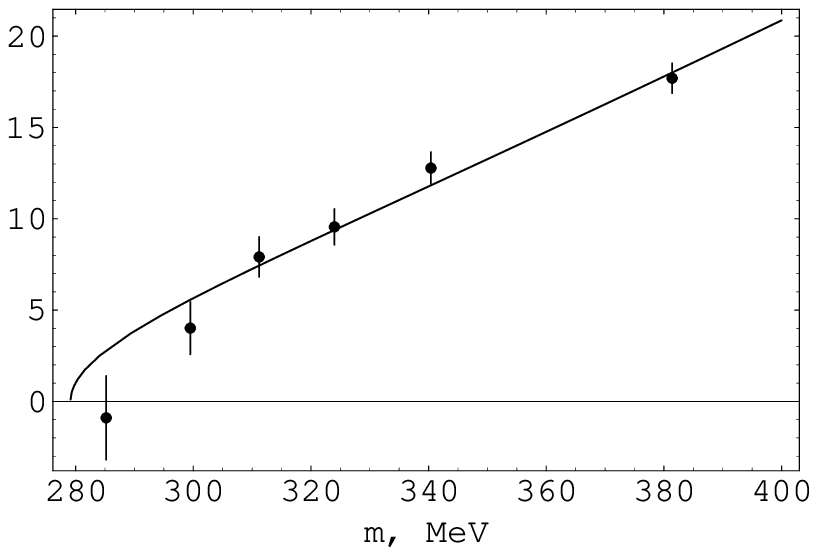}$}\\ (a)&(b)
\end{tabular}
\end{center}
\caption{The comparison of the experimental data on $\delta_0^0$
\cite{scatbnl} and the obtained curve is shown: a) Fit 1, b) Fit
2.} \label{fig4}
\end{figure}

\begin{figure}[h]
\begin{center}
\begin{tabular}{ccc}
\includegraphics[width=8cm]{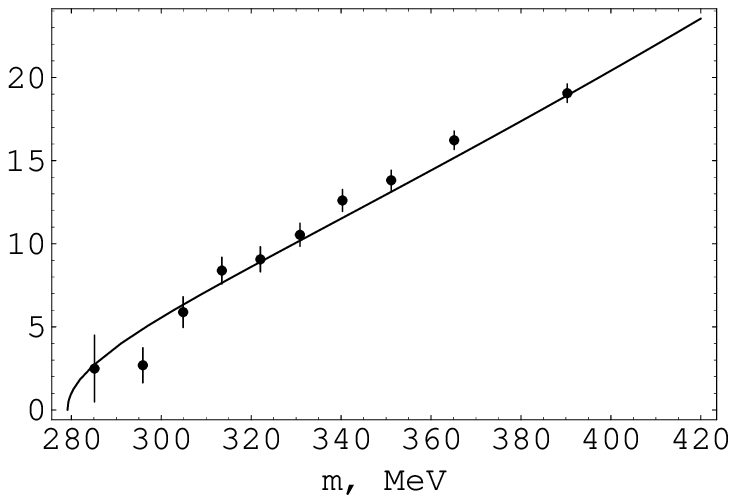}& \raisebox{-1mm}{$\includegraphics[width=8cm]{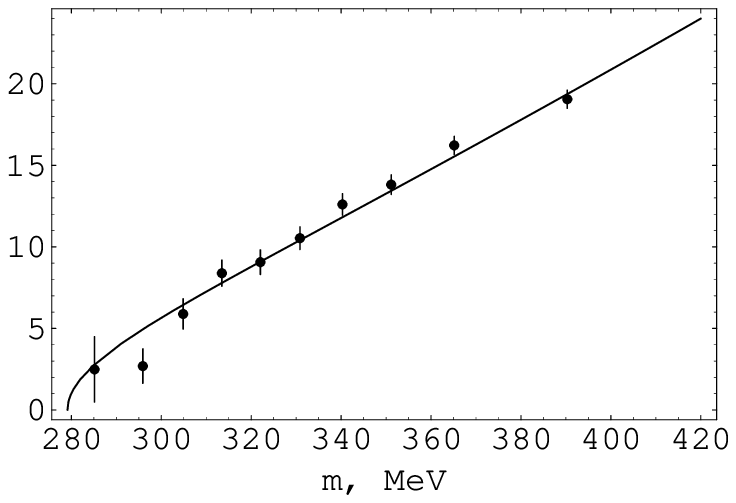}$}\\ (a)&(b)
\end{tabular}
\end{center}
\caption{The comparison of the experimental data on $\delta_0^0$
\cite{na48} and the obtained curve is shown: a) Fit 1, b) Fit 2.}
\label{fig5}
\end{figure}

\begin{figure}[h]
\begin{center}
\begin{tabular}{ccc}
\includegraphics[width=8cm]{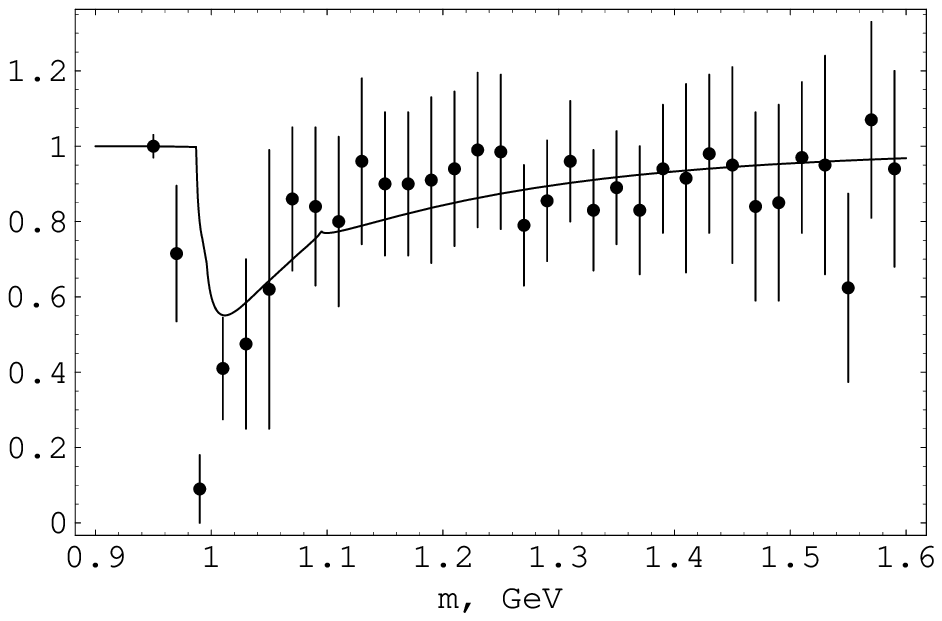}& \raisebox{-1mm}{$\includegraphics[width=8cm]{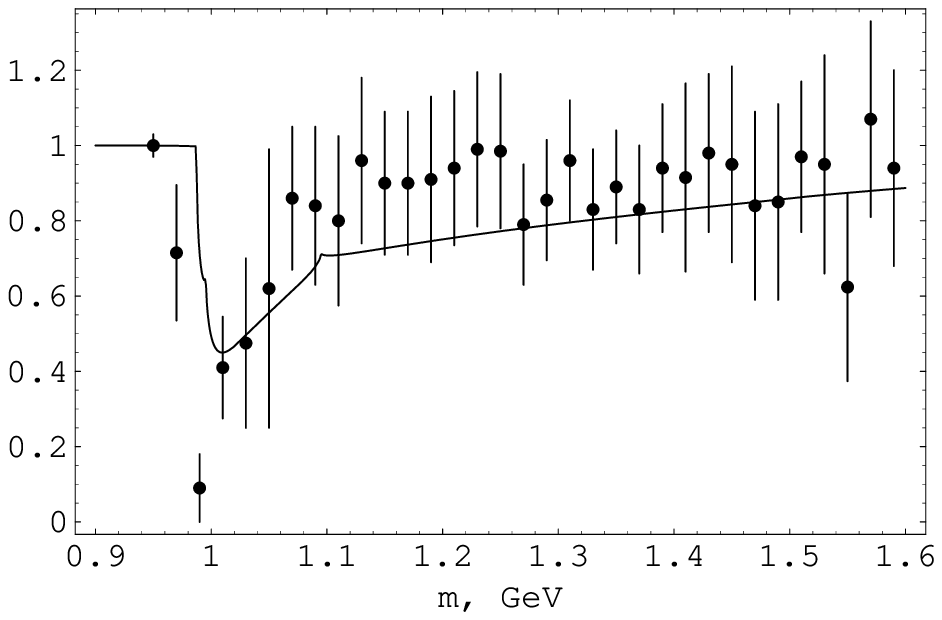}$}\\ (a)&(b)
\end{tabular}
\end{center}
\caption{The inelasticity $\eta^0_0$ is shown: a) Fit 1, b) Fit
2.} \label{fig6}
\end{figure}

\begin{figure}[h]
\begin{center}
\begin{tabular}{ccc}
\includegraphics[width=8cm]{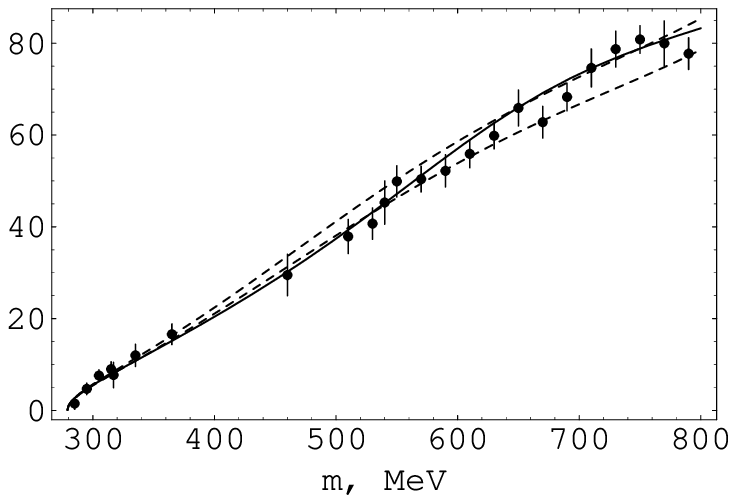}& \raisebox{-1mm}{$\includegraphics[width=8cm]{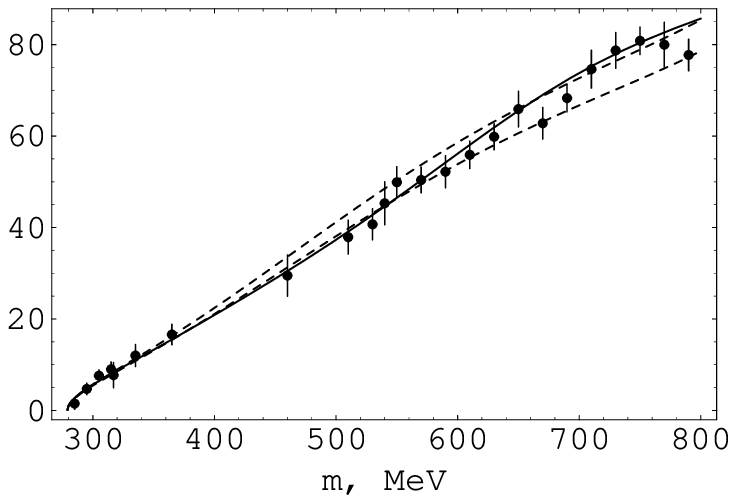}$}\\ (a)&(b)
\end{tabular}
\end{center}
\caption{The phase $\delta_0^0$ of the $\pi\pi$ scattering is
shown. The solid line is our description, dashed lines mark
borders of the corridor \cite{sigmaPole}, and points are
experimental data: a) Fit 1, b) Fit 2.} \label{fig7}
\end{figure}

\begin{figure}[h]
\begin{center}
\begin{tabular}{ccc}
\includegraphics[width=8cm]{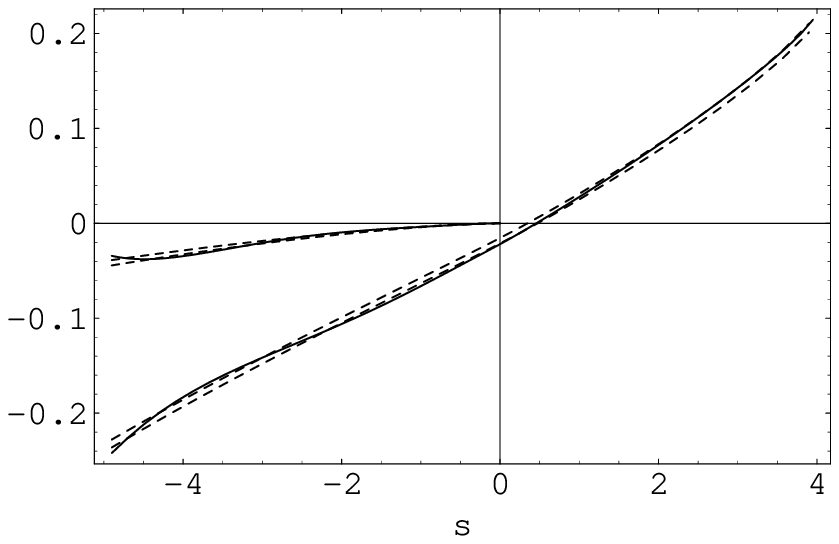}& \raisebox{-1mm}{$\includegraphics[width=8cm]{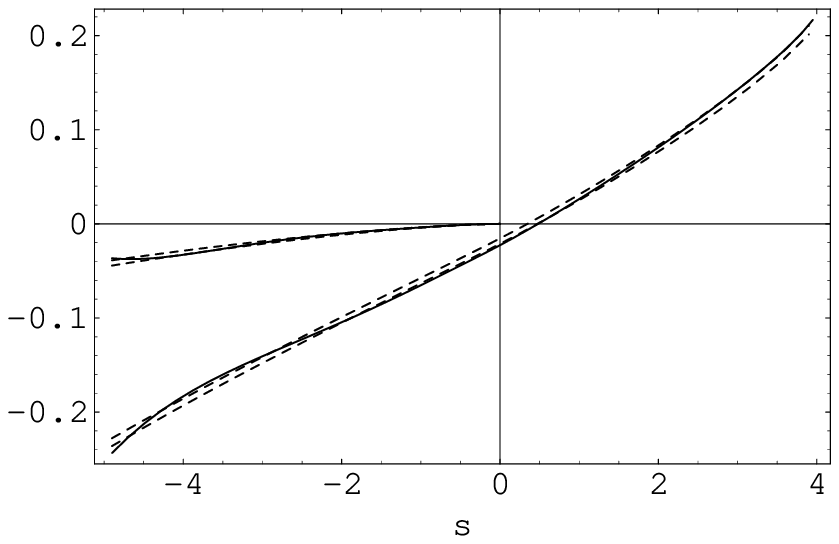}$}\\ (a)&(b)
\end{tabular}
\end{center}
\caption{The real and the imaginary parts of the amplitude $T^0_0$
of the $\pi\pi$ scattering ($s$ in units of $m_\pi^2$) are shown.
Solid lines show our description, dashed lines mark borders of the
real part corridor and the imaginary part for $s < 0$
\cite{sigmaPole}: a) Fit 1; b) Fit 2. } \label{fig8}
\end{figure}

\begin{figure}[h]
\begin{center}
\begin{tabular}{ccc}
\includegraphics[width=5cm]{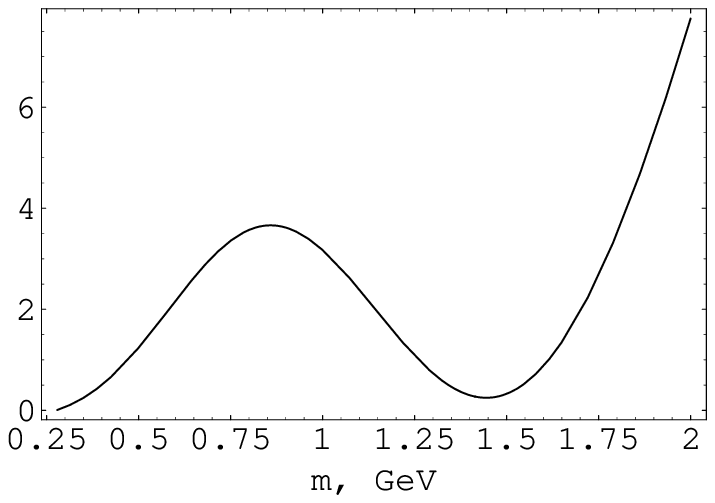}& \raisebox{-1mm}{$\includegraphics[width=5cm]{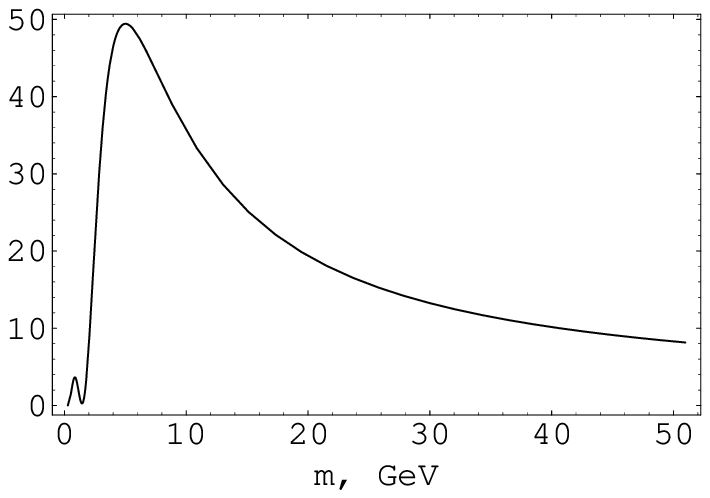}$}&
\raisebox{-1mm}{$\includegraphics[width=5cm]{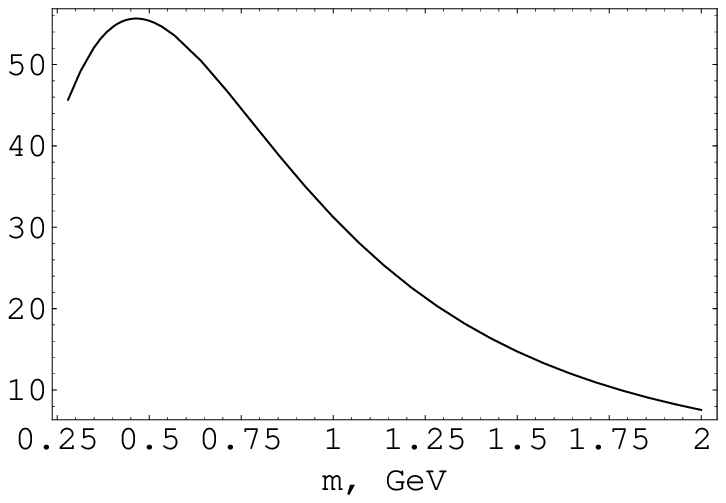}$}\\
(a)&(b)&(c)
\end{tabular}
\end{center}
\caption{Kernels $K_1(m^2)$ and $K_2(m^2)$ for Fit 1 are shown: a)
$K_1(m^2)$ below $2$ GeV. The minimum near $1.4$ GeV is $0.25$. b)
$K_1(m^2)$ up to $50$ GeV, then it asymptotically tends to $1$. c)
$K_2(m^2)$ up to $2$ GeV, then it asymptotically tends to zero. }
\label{fig9}
\end{figure}

\begin{figure}[h]
\begin{center}
\begin{tabular}{ccc}
\includegraphics[width=8cm]{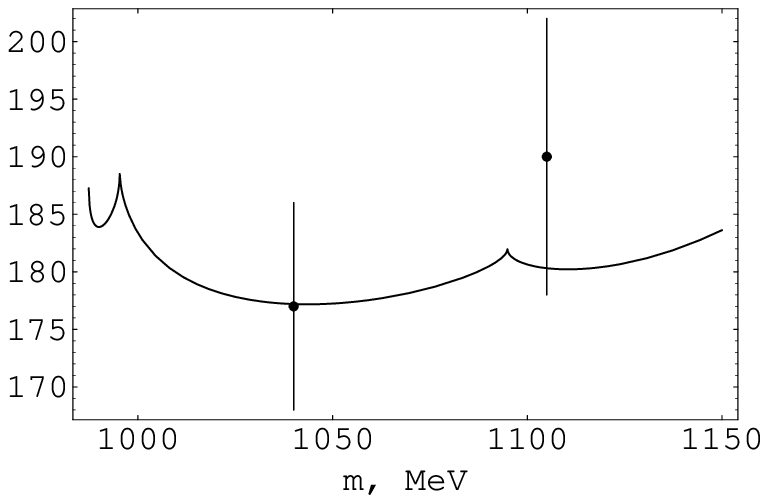}& \raisebox{-1mm}{$\includegraphics[width=8cm]{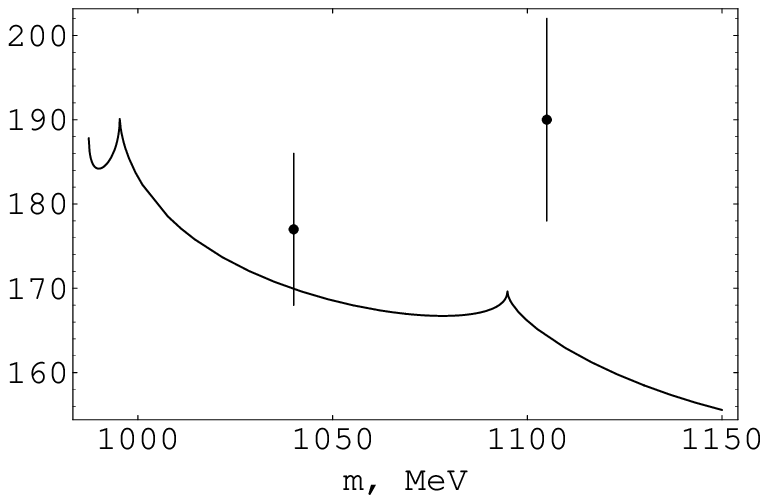}$}\\ (a)&(b)
\end{tabular}
\end{center}
\caption{The phase $\delta^{\pi K}$ of the $\pi\pi\to K\bar K $
scattering is shown: a) Fit 1; b) Fit 2.} \label{fig10}
\end{figure}

\begin{figure}[h]
\begin{center}
\begin{tabular}{ccc}
\includegraphics[width=8cm]{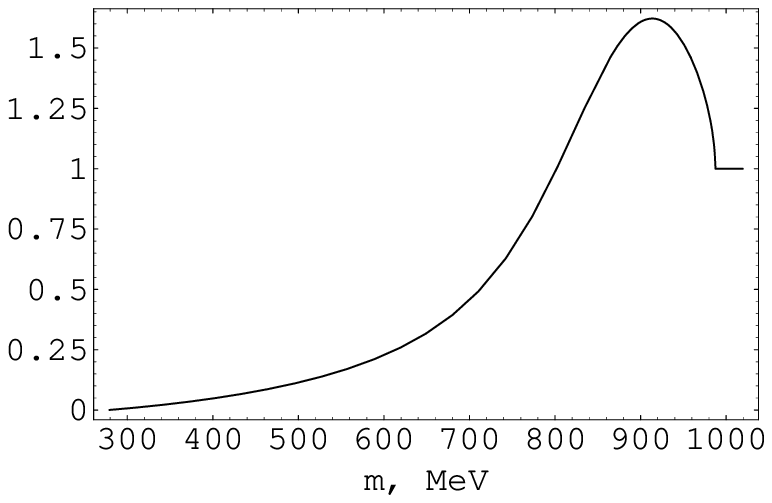}& \raisebox{-1mm}{$\includegraphics[width=8cm]{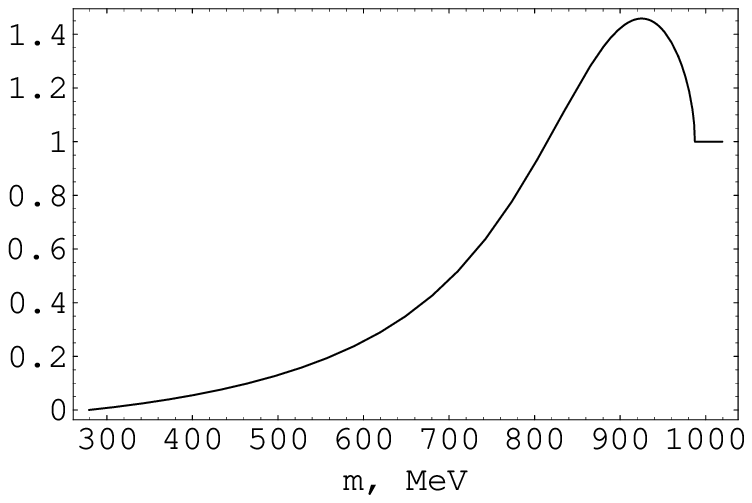}$}\\ (a)&(b)
\end{tabular}
\end{center}
\caption{The $|P_K(m)|^2$ is shown, see Eq. (7): a) Fit 1; b) Fit
2.} \label{fig11}
\end{figure}

\begin{figure}[h]
\begin{center}
\includegraphics[width=12cm]{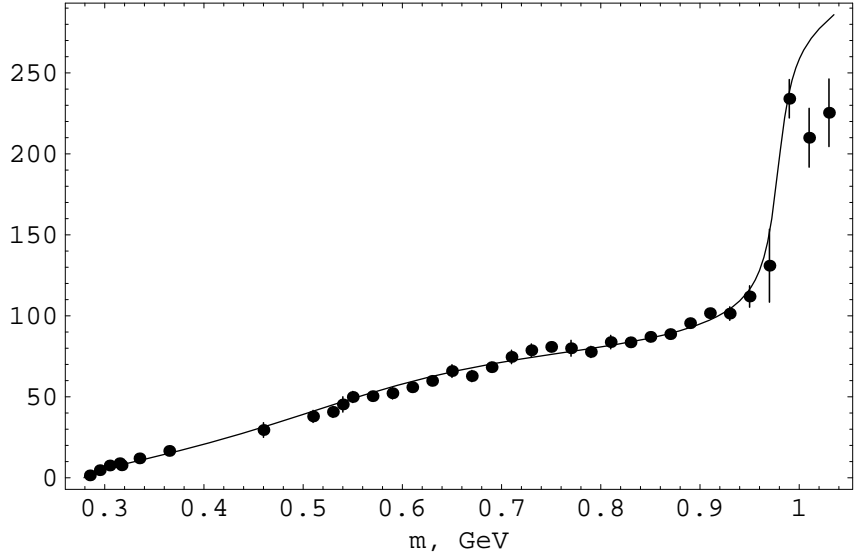}
\end{center}
\caption{The phase $\delta_0^0$ of the $\pi\pi$ scattering, Fit 3
and the experimental data are shown.} \label{fig12}
\end{figure}

\begin{figure}[h]
\begin{center}
\begin{tabular}{ccc}
\includegraphics[width=8cm]{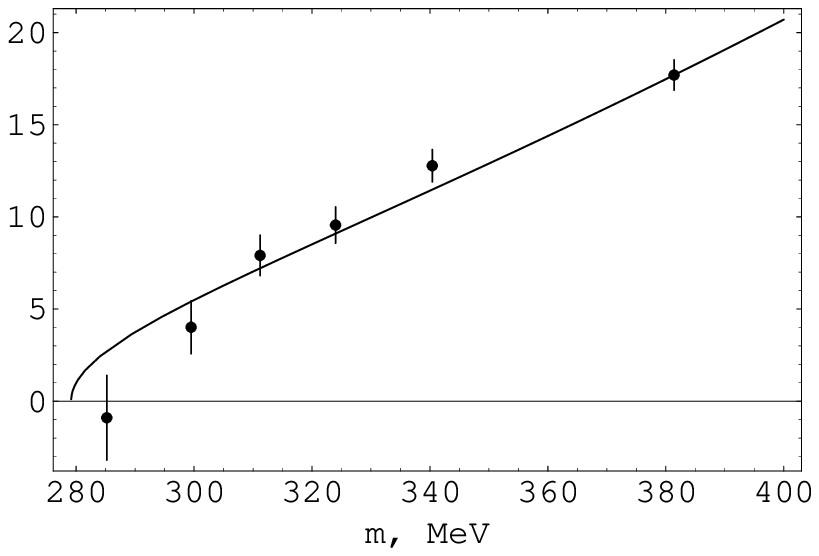}& \raisebox{-1mm}{$\includegraphics[width=8cm]{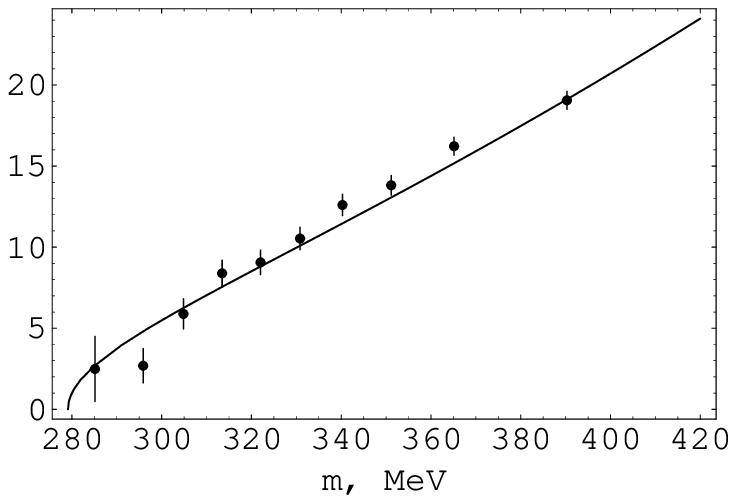}$}\\
(a)&(b)
\end{tabular}
\end{center}
\caption{The phase $\delta_0^0$ of the $\pi\pi$ scattering, Fit 3
is shown. The comparison with the data is available from a)
\cite{scatbnl}, b) \cite{na48}.} \label{fig13}
\end{figure}

\begin{figure}[h]
\begin{center}
\begin{tabular}{ccc}
\includegraphics[width=8cm]{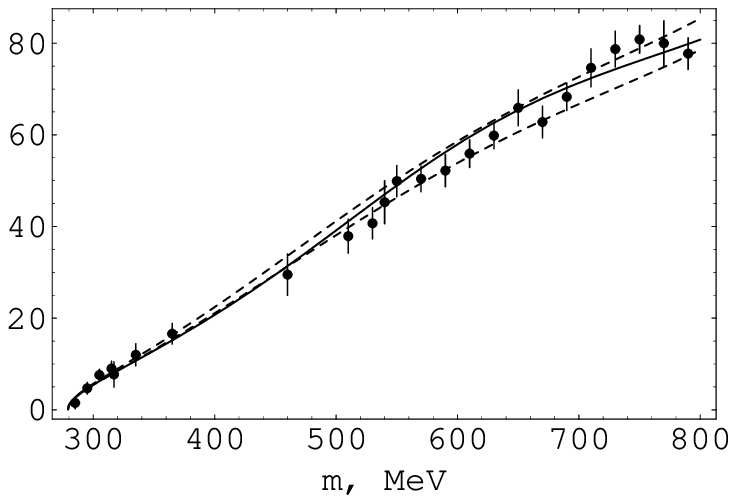}& \raisebox{-1mm}{$\includegraphics[width=8cm]{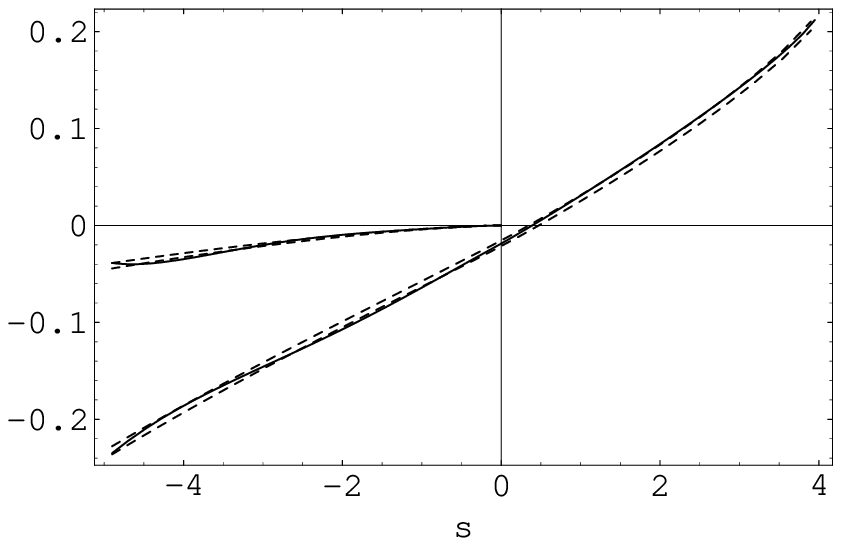}$}\\ (a)&(b)
\end{tabular}
\end{center}
\caption{a) The phase $\delta_0^0$ of the $\pi\pi$ scattering, Fit
3 is shown. The solid line is our description, dashed lines mark
borders of the corridor \cite{sigmaPole}, and points are
experimental data. b) The real and the imaginary parts of the
amplitude $T^0_0$ of the $\pi\pi$ scattering ($s$ in units of
$m_\pi^2$) are shown. The solid lines correspond to Fit 3, dashed
lines mark borders of the real part corridor and the imaginary
part for $s < 0$ \cite{sigmaPole}. } \label{fig14}
\end{figure}

Note it would be naive to treat the poles in the background as
resonances [$f_0(1370)$, for example] because in our approach to
consider additional resonances one should extend the matrix of the
inverse propagators, etc.

\section{Conclusion}
\label{sc}

Thus, the background phase (\ref{phB}) allows us to obtain proper
analytical features of the $\pi\pi$ scattering amplitude, link
results of \cite{sigmaPole} with properties of light scalars
simultaneously describing experimental data. The obtained
description is in agreement with the scenario based on the
four-quark model. The main signatures of this scenario are the
weak coupling of the $\sigma(600)$ meson   with the $K\bar K$
channel compared to the $\pi\pi $ one and   the weak coupling of
the $f_0(980)$ meson with the $\pi\pi$ channel  compared to the
$K\bar K$ one, see Table I, that results in the weak
$\sigma(600)-f_0(980)$ mixing  \cite{our_f0}. The ratios
$(g_{\sigma K^+K^-}/g_{\sigma \pi^+\pi^-})^2\approx 0.05-0.1$ and
$(g_{f_0\pi^+\pi^-}/g_{f_0 K^+K^-})^2\approx 0.15-0.18$, see Table
I, indicate roughly that the 90 percentage of  $\sigma (600)$ is
$\bar u\bar d u d$ and the 80 percentage of $f_0(980)$ is $\bar s
\bar d ds$.

Resonance masses and widths $m_R$ and $\Gamma_R(m_R)$ in our
formulas (which may be called "Breit-Wigner" masses and widths)
have clear physical meaning, in contrast to the resonance poles in
the complex plane. At first, what sheet of the complex plane
should be considered?

For $\sigma(600)$ it is natural to consider the first line of
Table IV [at any rate, it would be correct for very narrow
$\sigma(600)$]. The obtained pole positions in this case do not
agree with the pole position obtained in Ref. \cite{sigmaPole},
see Eq. (\ref{poleSigma}). Note that the $\sigma (600)$ pole
position is dictated by the $\sigma (600)$ propagator in our case,
because the $\sigma (600)-f_0(980)$ mixing is small. Providing the
pole position (\ref{poleSigma}) and taking into account only the
$\pi\pi$ channel in the propagator, we can determine $\sigma
(600)$ mass and coupling to the $\pi\pi$ channel, and the obtained
values contradict the K\"allen-Lehmann representation, see
\cite{ourProp}. Taking into account additional channels we may
fulfill the K\"allen-Lehmann representation, but the region of
permitted $\sigma (600)$ parameters do not allow us to describe
experimental data in the current model.

Note that the Roy equations are approximate, they take into
account only the $\pi\pi$ channel. This can lead to a different
analytical continuation and, hence, explain deviation of the
$\sigma (600)$ pole position, compare Fit 3 with Fits 1 and 2 in
Table I \cite{diffPropagators}.

The current activity, aiming extremely precise determination of
the $\sigma(600)$ pole position, has taken the forms of the
Swift's grotesque. Really, the residue of the $\sigma$ pole can
not be connected to coupling constant in the Hermitian (or
quasi-Hermitian) Hamiltonian, see Ref. \cite{annshgn-07}, for it
has a large imaginary part and this pole can not be interpreted as
a physical state for its huge width.

The futility of the approach that is based on the poles treatment
may be additionally illustrated by Fit 2. As seen on line 1 of
Table V, the real part of the $f_0(980)$ pole $Re M_{f_0}$ on the
II sheet of the $T^0_0$ exceeds the $K^+K^-$ threshold ($987.4$
MeV), it means that $Im M_{f_0}$ equals to
$-\bigg(\Gamma(f_0(980)\to\pi\pi)-\Gamma(f_0(980)\to
K^+K^-)\bigg)/2$, which is physically meaningless. In this case we
should take $\Pi^{K^+ K^-}$ from the second sheet, this gives the
pole at $M_{f_0}=989.6-168.7\,i$ MeV, with $Re M_{f_0}$ between
the $K^+K^-$ and $K^0\bar{K^0}$ thresholds again. As we work on
the $s$ plane, we should consider not $M_{f_0}$, but $M_{f_0}^2 =
(0.951 - 0.334\,i)$ GeV$^2$. So, we have the pole with a real part
below the $K^+K^-$ and $K^0\bar{K^0}$ thresholds and an imaginary
part dictated by analytical continuation of the kaon polarization
operators.

To reduce an effect of heavier isosinglet scalars we restrict
ourselves to the analysis of the mass region $m<1.2$ GeV. As to
mixing light and heavier isosinglet scalars, this question could
not be resolved once and for all at present, in particular,
because their properties are not well established up to now. A
preliminary consideration was carried out in Ref. \cite{joe},
where, in particular, it was shown that the mixing could affect
the mass difference of the isoscalar and isovector.

The factor $|P_K(s)|^2$ modifying the $|T(\pi\pi\to K\bar K)|^2$,
see Eqs. (7) and (\ref{fK}), is shown on Fig. \ref{fig11}. This
factor does not change the kaon loop model radically, but helps to
fulfill the requirement (\ref{kcondition}) and to improve the data
description. The influence of this factor may be reduced in order
to use a more skillful form than Eq. (\ref{fK}) for it.

New precise experimental data are needed for the investigation of
light scalars. The elucidation of the situation, a contraction of
the possible variants or even the selection of the unique variant,
requires considerable effort. The new precise experiment on
$\pi\pi\to K\bar K$ would give the crucial information about the
inelasticity $\eta^0_0$ and about the phase $\delta_B^{K\bar
K}(m)$ near the $K\bar K$ threshold. The forthcoming precise
experiment in KLOE on the $\phi\to\pi^0\pi^0\gamma $ decay will
also help to judge this phase in an indirect way. The precise
measurement of the inelasticity $\eta^0_0$ near 1 GeV in
$\pi\pi\to\pi\pi $ would also be very important.

It is of interest to update our analysis of the $\phi\to
a_0(980)\gamma \to\eta\pi^0\gamma$ decay \cite{our_a0} and the
$\gamma\gamma \to a_0(980)\to\pi^0\eta$ description
\cite{gg_to_aeta} in this analytical approach. Probably, such an
approach would also be useful for the $\kappa(900)$ meson
investigation in the $\pi K$ channel.

\section{Acknowledgements}
We thank very much H. Leutwyler for providing numerical values of
the $T^0_0(s)$ on the real axis, obtained in Ref.
\cite{sigmaPole}, useful discussions, and kind contacts. This work
was supported in part by RFBR, Grant No 10-02-00016.


\begin{thebibliography}{99}
\bibitem{TbackIs0}
We met the opinion that the representation (\ref{SSrepres})
violates analiticity, because the amplitude

$$T_{\pi\pi}=T_{res}+T_{back}+2i\rho_{\pi\pi} T_{res}T_{back}$$

\noindent becomes infinite at $s\to 0$ since
$\rho_{\pi\pi}\to\infty$ in this case
($\rho_{\pi\pi}=\sqrt{1-4m_\pi^2/s}$). This obstacle may be
overcome if $T_{res}$ or $T_{back}$ is equal to zero at $s=0$. As
in Ref. \cite{our_f0}, we choose here the second variant,
$T_{back}(0)=0$.
\bibitem{smallRhoPi}
$F_b$ may be considerably less than $1$ due to the $\pi^0\pi^0$
final state interaction.
\bibitem{AdlerZero}
Note that, in general, the existence of Adler zero in partial
amplitudes is not obligatory, see APPENDIX II in Ref.
\cite{our_f0}.
\bibitem{theSameData}
The same data was used in \cite{our_f0}, but only up to $1.1$ GeV.
We add the data up to $1.2$ GeV to provide better behaviour of our
$\delta^0_0$ and inelasticity, considering influence of
high-energy resonances smaller than experimental errors.
\bibitem{equalConstants}
These constants are equal in the naive four-quark model.
\bibitem{diffPropagators}
In addition, propagators (\ref{propagator}) are approximate, and
their small difference with "real" propagators in the physical
region could be important in the complex $s$ plane.
\end{thebibliography}
\end{document}